\documentclass[universe,review,accept,moreauthors,pdftex]{Definitions/mdpi}

\firstpage{1} 
\pubvolume{8}
\issuenum{1}
\articlenumber{0}
\pubyear{2021}
\copyrightyear{2021}
\datereceived{} 
\dateaccepted{} 
\datepublished{} 
\hreflink{https://doi.org/} 

\usepackage{amssymb,amsthm,mathtools}
\usepackage{amsmath, textcomp}
\usepackage[labelformat=simple]{subcaption} 

\DeclareCaptionLabelFormat{subcaptionlabel}{\normalfont(\textbf{#2}\normalfont)}
\usepackage{xfrac}    
\usepackage{xcolor}

\Title{Design of New Resonant Haloscopes in the Search for the Dark~Matter Axion: A Review of the First Steps in the RADES~Collaboration}

\TitleCitation{Design of New Resonant Haloscopes in the Search for the Dark Matter Axion: A Review of the First Steps in the RADES Collaboration}

\Author{\textls[-10]{Alejandro D\'iaz-Morcillo $^{1,}$*, Jos\'e Mar\'ia Garc\'ia~Barcel\'o $^{1}$, Antonio J.~Lozano-Guerrero $^{1}$, Pablo Navarro $^{1}$, Benito Gimeno $^{2}$, Sergio Arguedas~Cuendis $^{3}$,} Alejandro \'Alvarez~Melc\'on $^{1}$,  Cristian Cogollos $^{4}$, Sergio Calatroni $^{3}$, Babette D\"obrich $^{3}$, Juan Daniel Gallego $^{5}$, Jessica Golm $^{3,6}$, Igor Garc\'ia Irastorza $^{7}$, Chloe Malbrunot $^{3}$, Jordi Miralda-Escud\'e $^{4,8}$, Carlos Pe\~na Garay $^{9,10}$, Javier Redondo $^{7,11}$ and Walter Wuensch $^{3}$}

\AuthorNames{Alejandro D\'iaz-Morcillo, Jos\'e Mar\'ia Garc\'ia~Barcel\'o, Antonio J. Lozano-Guerrero, Pablo Navarro, Benito Gimeno, Sergio Arguedas~Cuendis, Alejandro \'Alvarez~Melc\'on,  Cristian Cogollos, Sergio Calatroni, Babette D\"obrich, Juan Daniel Gallego, Jessica Golm, Igor Garc\'ia Irastorza, Chloe Malbrunot, Jordi Miralda-Escud\'e, Carlos Pe\~na Garay, Javier Redondo and Walter Wuensch}

\AuthorCitation{D\'iaz-Morcillo, A.; Garc\'ia~Barcel\'o, J.M.; Lozano-Guerrero, A.J.; Navarro, P.; Gimeno, B.; Arguedas~Cuendis, S.; \'Alvarez~Melc\'on, A.; Cogollos, C.; Calatroni, S. and D\"obrich, B.; et al.}

\address{%
$^{1}$ \quad Department of Information and Communications Technologies, Universidad Politécnica de Cartagena, 30202~Cartagena, Spain; josemaria.garcia@upct.es (J.M.G.B.); antonio.lozano@upct.es (A.J.L.-G.); pablonm.ct.94@gmail.com (P.N.); alejandro.alvarez@upct.es (A.\'A.M.)\\
$^{2}$ \quad Instituto de F\'isica Corpuscular (IFIC), CSIC-University of Valencia, 46071 Valencia, Spain; benito.gimeno@uv.es \\
$^{3}$ \quad European Organization for Nuclear Research (CERN), 1211 Geneva 23, Switzerland; sergio.arguedas.cuendis@cern.ch (S.A.C.); Sergio.Calatroni@cern.ch (S.C.); babette.dobrich@cern.ch (B.D.); jessica.golm@cern.ch (J.G.); chloe.m@cern.ch (C.M.); Walter.Wuensch@cern.ch (W.W.)\\ 
$^{4}$ \quad Institut de Ci\`encies del Cosmos, Universitat de Barcelona
(UB-IEEC), 08028 Barcelona, Spain; ccogollos@icc.ub.edu (C.C.); jmiralda@fqa.ub.edu (J.M.-E.)\\ 
$^{5}$ \quad Yebes Observatory, National Centre for Radioastronomy Technology and Geospace Applications, 19080~Guadalajara, Spain; jd.gallego@oan.es\\
$^{6}$ \quad Institute for Optics and Quantum Electronics, Friedrich Schiller University Jena, 07743 Jena, Germany\\ 
$^{7}$ \quad Center for Astroparticle and High Energy Physics (CAPA), Departamento de F\'isica Te\'orica, Universidad de Zaragoza, 50009 Zaragoza, Spain; Igor.Irastorza@cern.ch (I.G.I.); jredondo@unizar.es (J.R.)\\ 
$^{8}$ \quad Instituci\'o Catalana de Recerca i Estudis Avan\c cats, 08028 Barcelona, Spain\\ 
$^{9}$ \quad I2SysBio, CSIC-University of Valencia, 46071 Valencia, Spain; cpenya@lsc-canfranc.es\\
$^{10}$\quad Laboratorio Subterr\'aneo de Canfranc, 22880 Estaci\'on de Canfranc, Spain\\ 
$^{11}$\quad Max-Planck-Institut f\"ur Physik (Werner-Heisenberg-Institut), 80805 M\"unchen, Germany}

\corres{Correspondence: alejandro.diaz@upct.es}

\abstract{With the increasing interest in dark matter axion detection through haloscopes, in which different international groups are currently involved, the RADES group was established in 2016 with the goal of developing very sensitive detection systems to be operated in dipole magnets. This review deals with the work developed by this collaboration during its first five years: from the first designs---based on the multi-cavity concept, aiming to increase the haloscope volume, and thereby improve sensitivity---to their evolution, data acquisition design, and finally, the first experimental run. Moreover, the envisaged work within RADES for both dipole and solenoid magnets in the short and medium term is also presented.}


\begin{document}


\section{Introduction}

Since Peccei and Queen established a mechanism to solve the strong Charge--Parity {(CP)} problem \cite{PecceiQuinn1977Jun,PecceiQuinn1977Sep}, and subsequently Weinberg \cite{Weinberg1978} and Wilczek \cite{Wilczek1978} predicted the existence of a new particle, the axion, the defining and execution of different experimental setups in order to detect this proposed particle have gained an increasing interest. {In addition, soon after their proposal, axions were acknowledged to be ideal dark matter (DM) candidates, due to non-thermal production channels in the early Universe that are directly expected from the very Peccei--Quinn mechanism~\cite{Preskill:1982cy,Abbott1983,Dine1983}. More recently, due in part to the persistently negative outcomes of the many worldwide efforts to detect weakly interacting massive particles (WIMPs), the favorite DM candidate for the last three decades, the axion DM hypothesis, has been attracting more interest in the experimental community.}

{Apart from their gravitational interaction with normal matter, axions are also expected to have a weak electromagnetic (photons) and Dirac fermionic (e.g., nucleons or electrons) coupling. In this work, we focus on the axion coupling into two photons given by
\begin{equation}
    \label{Lagr}
    \mathcal{L}_{a\gamma\gamma} \propto g_{a\gamma} a \vec{E}\cdot\vec{B}_{e},
\end{equation}
where $g_{a\gamma} =  \alpha g_{\gamma}/2\pi f_a$ is the axion--photon coupling coefficient, $\alpha$ the fine structure constant, $f_a$ the axion decay constant, and $g_{\gamma}$ a model dependent dimensionless coupling constant. $\vec{E}\cdot\vec{B}_{e}$ is the scalar product of the electric field and an externally applied magnetic field, and $a$ the axion field.}

In the last thirty years, the search {based on axion--photon coupling} has been developed following three different detection techniques, depending on the sources of these particles and their expected masses. Two of them, based on the inverse Primakoff \mbox{effect \cite{Primakoff1951}}, the so called axion helioscopes and haloscopes, were proposed by Sikivie \cite{Sikivie1983} in the mid-80s. The other one, known as Light Shining through Walls (LSW) {\cite{VanBibber(1987),Fukuda(1996),Sikivie(2007)}} is based on both direct and inverse Primakoff effects. {Helioscopes and haloscopes detect axions from external sources (the sun or relic axions), whereas LSW generates axions by itself.}

{The LSW uses two Fabry--Perot cavities separated by an optical barrier. Axions are generated in the first cavity and detected in the second. Both generation and detection are through the coupling between photons and axions.}

Helioscope experiments look for axions produced in the Sun through the Primakoff effect from plasma photons in the solar inner layers. After initial experiments at Brookhaven National Laboratory (BNL) in 1992 \cite{Lazarus1992} and in Japan in the SUMICO experiment \cite{Inoue2002,Moriyama1998,Inoue2008}, the third-generation CAST experiment at CERN obtained the best results in this kind of detection up to now, with limits for axion--photon coupling <$6.6\times10^{-11}$ ${\rm GeV}^{-1}$ in the mass range <0.02 eV \cite{CAST:2017uph}
, and <$2.3\times10^{-10}$ ${\rm GeV}^{-1}$ for 0.02 eV < $m_a$ < 0.64 eV \cite{Arik2011}. The future of solar axion searching will star IAXO \cite{Armengaud2019}. 

{The quest for dark matter in the form of axions brings up the idea of a haloscope experiment. In 2016, the RADES (Relic Axion Detector Exploratory Setup)} collaboration started, joining research groups from universities and research institutes of Spain and researchers at CERN. A detailed state-of-the-art report of the different axion detection experiments, finished or under development and execution, can be found in \cite{Graham2015}. Updated {and comprehensive} reviews can be found in {\cite{Irastorza2018,DiLuzio:2020wdo,Sikivie2021}}. The aim of the collaboration was the detection of the axion in our galactic halo, which is, in fact, a candidate for constituting dark \mbox{matter \cite{Abbott1983,Dine1983}}. Therefore, the experiment should be clearly based on a haloscope setup. Haloscope experiments intend to detect axions in the galactic halo by promoting the axion--photon conversion under a strong magnetic field. The intensity of the electromagnetic field is boosted when this conversion occurs in a resonant device, such as a microwave resonant cavity. Additionally, due to the expected extremely low axion--photon coupling, and thus the electromagnetic power to be detected, it is mandatory to work in cryogenic conditions in order to minimize the thermal noise, and thus to increase the signal to noise ratio ($\sfrac{S}{N}$).

At the beginning of the RADES collaboration, the successful experiment ADMX had already completed the exploration in the 1.9--3.65 $\upmu$eV mass range (460--890 MHz) with exclusion limits that reached those of the KSVZ model \cite{Stern2016}. It benefited from the experience of two pioneering experiments at BNL \cite{Wuensch1989} and the University of Florida \cite{Hagmann1990}, respectively. The prospects for the new version of the experiment (ADMX-Gen2) reach even DFSZ sensitivity in the 1.4--25 $\upmu$eV mass range \cite{Stern2016}.

Since the RADES project began, the research with resonant cavity haloscopes has greatly increased, with a large number of new cavity concepts, such as the KLASH proposal \cite{Alesini2017} at lower masses than ADMX; and HAYSTACK \cite{Kenany2017}, ORGAN \cite{McAllister2017}, QUAX \cite{Alesini2021}, and CAPP experiments \cite{Jeong2018} at higher ones. RADES \cite{Alvarez2018} is among this latter group.

\section{General Principles in Resonant Haloscope Design}

As described in \cite{Alvarez2018} for the RADES case, a resonant haloscope experiment consists of a resonant cavity immersed in a high and static magnetic field at very low temperatures (few K or even mK), and connected to a receiver or data acquisition system (DAQ) which must amplify/filter/down-convert the detected signal without introducing too much noise and perform the A/D conversion and subsequent fast Fourier transform (FFT) for a proper offline data analysis.

We can summarize the main goals in the design of an axion detection experiment in (1) maximizing the electromagnetic power detected from the axion--photon coupling, (2)~optimizing the sensitivity of the experiment, and (3) maximizing the scanning rate along the mass range of interest. It is of key interest, thus, to know the relations of these three parameters with the experimental variables.

\subsection{Detected Power}

In a resonant cavity haloscope experiment, the  power generated in the axion--photon conversion can be obtained as \cite{Younggeun2019}
\begin{equation}
    P_g = g^2_{a\gamma}\frac{\rho_a}{m_a}B_e^2CVmin(Q_c,Q_a),
\end{equation}
or more accurately \cite{Dongok2020},
\begin{equation}
    P_g = g^2_{a\gamma}\frac{\rho_a}{m_a}B_e^2CV\frac{Q_cQ_a}{Q_c+Q_a}
\end{equation}
where $g_{a\gamma}$  is the unknown axion--photon coupling {in (\ref{Lagr})}; $\rho_a$ the axion dark matter density; $m_a$ the axion mass; $B_e$ the static magnetic field; $V$ the cavity volume; $Q_a$ and $Q_c$ the quality factors of the axion and cavity resonances, respectively; and $C$ the form or geometric factor, a normalized parameter that quantifies the coupling between the static magnetic field $\vec{B}_e$ and the dynamic electric field $\vec{E}$ induced by the axion--photon conversion:
\begin{equation}
\label{C}
    C = \frac{|\int_V \vec{E}\cdot\vec{B}_e \, dV|^2}{\int_V ||\vec{B}_e||^2 \,dV \, \int_V \varepsilon_r ||\vec{E}||^2 \, dV}
\end{equation}
where simple bars $|\cdot|$ indicate the magnitude of a complex number, double bars $||\cdot||$ are the standard vector norm defined in the 3D complex vector space, and  $\varepsilon_r$ is the dielectric constant inside the cavity.

This generated power will be in part dissipated in the lossy material of the cavity (metallic walls if there are no lossy dielectric materials) and in part extracted from the cavity with a proper probe. This coupling, for frequencies below the millimetric wave range, will be normally a monopole or loop antenna, depending on the mode to be extracted, connected to a coaxial line which, in turn, transmits the signal to the receiver. The power detected at the coaxial port is
\begin{equation}
 \label{Pd1}
   P_d = \kappa g^2_{a\gamma}\frac{\rho_a}{m_a}B_e^2CV\frac{Q_lQ_a}{Q_l+Q_a}
\end{equation}
where $\kappa$ is the coupling factor at the extraction point, and $Q_l$ is the loaded quality factor, that is, the quality factor of the cavity taking into account the coupling effect. Maximum power transfer is obtained in a critical coupling regime, that is, for $\kappa=0.5$. Although critical coupling can be designed, it is very sensitive to slight mechanical or temperature changes. Therefore, it is very convenient to develop mechanical or electromagnetic systems for modifying the coupling in situ.

Moreover, if $Q_l$ is much lower than $Q_a$ ($\sim$$10^6$), (\ref{Pd1}) becomes
\begin{equation}
\label{Pd}
    P_d = \kappa g^2_{a\gamma}\frac{\rho_a}{m_a}B_e^2CVQ_l
\end{equation}

\subsection{Axion Coupling Sensitivity}

The complete experiment, using a source (resonant haloscope) and a receiver, has a signal to noise ratio given by Dicke's radiometer equation:
\begin{equation}
\label{fracSN}
   \frac{S}{N} = \frac{P_d}{P_N} = \frac{P_d}{k_BT_{sys}\sqrt{\frac{\Delta v}{\Delta t}}}
\end{equation}
where $k_B$ is the Boltzmann constant, $T_{sys}$ the noise temperature of the system, $\Delta t$ the observation time, and $\Delta v$ the reception bandwidth, which should coincide with the bandwidth of the axion resonance:
\begin{equation}
\label{Deltava}
    \Delta v_a = \frac{m_a}{Q_a}
\end{equation}

\textls[-15]{By introducing (\ref{Pd}) and (\ref{Deltava}) into (\ref{fracSN}) we can obtain the expression for the axion--photon~coupling:}
\begin{equation}
\label{ga}
    g_{a\gamma} = \left(\frac{\frac{S}{N}k_BT_{sys}}{\kappa\rho_aCVQ_l}\right)^{\frac{1}{2}}\frac{1}{B_e}\left(\frac{m_a^3}{Q_a\Delta t}\right)^{\frac{1}{4}}
\end{equation}

An alternative parameter to $g_{a\gamma}$ is the dimensionless axion--photon coupling $C_{a\gamma}$, related to $g_{a\gamma}$  through
\begin{equation}
\label{Ca}
    C_{a\gamma} = 5\times10^{15}\frac{g_{a\gamma}\left(GeV^{-1}\right)}{m_a\left(\mu eV\right)}
\end{equation}

Introducing (\ref{ga}) into (\ref{Ca}) yields
\begin{equation}
\label{Cav2}
    C_{a\gamma} = 5\times10^{15}\left(\frac{\frac{S}{N}k_BT_{sys}}{\kappa\rho_aCVQ_l}\right)^{\frac{1}{2}}\frac{1}{B_e}\left(\frac{1}{Q_am_a\Delta t}\right)^{\frac{1}{4}}
\end{equation}

As seen in (\ref{Pd1}), the detected power only depends on intrinsic parameters of the axion ($g_{a\gamma}$, $\rho_a$, $m_a$, and $Q_a$) and experimental parameters in the cavity subsystem ($B_e^2$, $C$, $V$, and $Q_l$), but the sensitivity also relies on receiver experimental parameters ($\frac{S}{N}$, $T_{sys}$, and $\Delta t$).

\subsection{Scanning Rate}

Despite its high-Q resonant behavior, the final aim of a cavity haloscope is exploring the largest possible mass range and doing it quickly. The mass sweeping can be performed by adequate moving mechanisms or tuning materials that modify the cavity's resonant frequency, and the scanning rate $\frac{dm_a}{dt}$ is usually referred to as a figure of merit of the haloscope.

For a measurement with the cavity resonance centered at $m_a$, the mass range explored is the bandwidth of the cavity:
\begin{equation}
    dm_a = \frac{m_a}{Q_l}
\end{equation}
and the time needed for that measurement in order to get a prescribed $\frac{S}{N}$ is, from (\ref{fracSN}),
\begin{equation}
    dt = \Delta v_a\left(\frac{\frac{S}{N}k_BT_{sys}}{P_d}\right)^2=\frac{m_a}{Q_a}\left(\frac{\frac{S}{N}k_BT_{sys}}{P_d}\right)^2
\end{equation}

Therefore,
\begin{equation}
\label{dmadt}
    \frac{dm_a}{dt} = \frac{Q_a}{Q_l}\left(\frac{P_d}{\frac{S}{N}k_BT_{sys}}\right)^2=
    Q_aQ_l\kappa^2g_{a\gamma}^4\left(\frac{\rho_a}{m_a}\right)^2B_e^4C^2V^2\left(\frac{S}{N}k_BT_{sys}\right)^{-2}
\end{equation}

\section{Motivation and Constraints of the RADES Project}

As a summary of the last section, the experimental parameters that optimize the detected power, the sensitivity, and the scanning rate can be identified in (\ref{Pd}), (\ref{Cav2}), and~(\ref{dmadt}), respectively. On the cavity side, it is evident that the most important parameter is the external magnetic field $B_e$, followed by the volume and the loaded quality factor of the cavity, the form factor for the axion-coupling electromagnetic mode, and the noise temperature of the system, which, due to the Friis formula \cite{Pozar1998}, is mainly determined by the physical temperature of the cavity and the noise temperature of the first amplifier.

A figure of merit for the magnet in this kind of experiment is the magnetic volume $B_e^2V$. However, it is usually difficult to design a haloscope that fills completely the available volume, so the potential of the magnet is underused. This was the  main motivation of the RADES project. Taking into account the upper limit for the ADMX prospects in the next few years \cite{Stern2016}, RADES aimed to explore higher masses, but these lead to smaller resonant cavities, since the resonant frequency depends, in general, inversely on the dimensions of the cavity.

For instance, continuing with the ADMX case, the resonant frequency of a circular cavity for TM modes is given by
\begin{equation}
    f_{TM_{nlq}} = \frac{c}{2\pi}\sqrt{\left(\frac{p_{nl}}{a}\right)^2+\left(\frac{q}{d}\right)^2}
\end{equation}
where $n$, $l$, and $q$ are the numbers of variations in azimuthal, radial, and axial coordinates, respectively; $c$ is the speed of light; $p_{nl}$ is the $l$-th root of the first-kind Bessel function of order $n$; and $a$ and $d$ are the radius and length of the cavity, respectively. For the $TM_{010}$ mode used in the ADMX experiment and an expected exploration mass of 30 $\upmu$eV (7.25 GHz) the cavity diameter should be 3.17 cm, which would underuse the available \mbox{60 cm bore}.

\begin{figure}[H]   
	        \includegraphics[width= 0.49\textwidth]{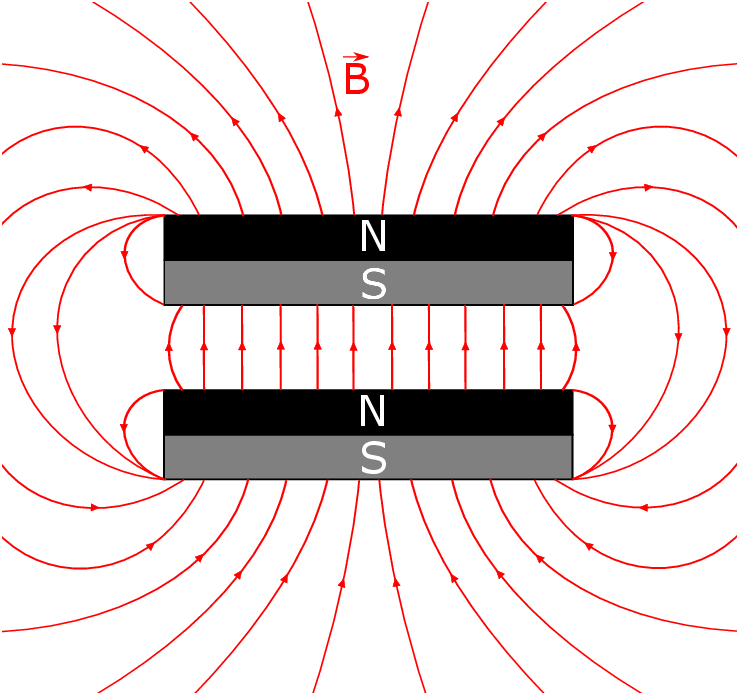}
	        \includegraphics[width=0.49\textwidth]{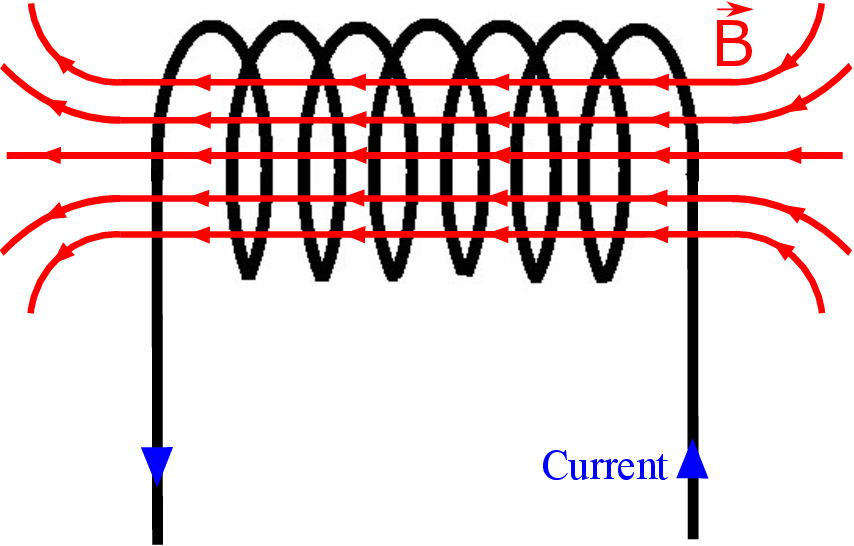}
		\caption{Dipole (\textbf{left}) and solenoid (\textbf{right}) magnets with transversal and axial magnetic \mbox{fields, respectively.}}
		\label{fig:Dipole&Solenoid}
\end{figure}

Most haloscope experiments, such as ADMX and HAYSTACK, use solenoid magnets. These magnets produce an axial ($\hat{z}$) magnetic field (Figure \ref{fig:Dipole&Solenoid} right), and therefore, the usual configuration for the cavity is a circular cavity with its axis parallel to the magnet bore axis. In this way, a TM mode, whose electric field is also axial, yields an optimal form factor. In the case of the RADES project, strong accelerator dipole magnets (\mbox{Figure \ref{fig:Dipole&Solenoid} left}) were available in the initial stages, concretely the CAST magnet, which yields a 8.8 T transverse ($\hat{y}$) magnetic field. A more convenient cavity for this kind of magnet is the rectangular cavity, where $TE_{m0p}$ modes have vertical polarization for the electric field.

For a rectangular cavity, the resonant frequencies for both TE and TM modes are \mbox{given by}
\begin{equation}
\label{fTE}
    f_{mnp} = \frac{c}{2}\sqrt{\left(\frac{m}{a}\right)^2+\left(\frac{n}{b}\right)^2+\left(\frac{p}{d}\right)^2}
\end{equation}
where $m$, $n$, and $p$ are the sinusoidal variations of the mode in the $x$, $y$, and $z$ dimensions, respectively; $a$ is the width, $b$ the height, and $d$ the length of the cavity.

In this case, for the mode with a maximum form factor, that is, the $TE_{101}$ mode, a mass goal of 30 $\upmu$eV, and to get the maximum volume ($a=b=d$), the cavity should be a hexahedron with $a=29.3$ mm. This fits with the 42.5 mm diameter of the bore in the CAST magnet \cite{Zioutas1999}, but the magnet is greatly underused if we compare the length of the cavity with the 9.3 m of the bore. Additionally, in order to separate degenerate modes, $a$ and $b$ must be sensibly different (let us assume $a\approx2b$). This reduces the cross-section of the cavity, and with it, its volume.

Taking this into account, and despite the fact that a rectangular cross-section is not the best way to maximize the use of a circular cross-section, the RADES project driving motivation was increasing the volume of the cavity in the axial direction without modifying the resonant frequency and other associated parameters, such as $Q_l$ and $C$.

\subsection{First Attempt: Increasing the Length of the Cavity}

Obviously, after a quick glimpse to (\ref{fTE}), the first idea for increasing the volume without modifying the resonant frequency is increasing the cavity length. For the $TE_{101}$ mode, (\ref{fTE}) is simplified to
\begin{equation}
\label{fTE101}
    f_{TE_{101}} = \frac{c}{2}\sqrt{\frac{1}{a^2}+\frac{1}{d^2}}
\end{equation}

Thus, if $a\ll d$ the length can be increased without modifying the resonant frequency, since $f_{TE_{101}}\approx\frac{c}{2a}$. This approach has a minor drawback: the reduction in $a$ (and, obviously in $b$ to keep $a\approx 2b$) reduces the cavity cross-section, and this contributes to a reduction of part of the volume gained with the larger $d$. Anyway, this reduction is small compared with the increase due to $d$ increment, as the following example shows.

Let us assume arbitrarily a goal exploration mass of 34.76 $\upmu$eV (8.4 GHz). For the $TE_{101}$ mode, fixing $a=d$ in (\ref{fTE101}) leads to $a=22.86$ mm, and fixing $d=10a$ yields \mbox{$a=17.95$ mm}. Increasing $d$ up to $500a$ just decreases the resonant frequency by less than $0.5\%$ and increases the volume by a factor $242$ with respect to the $a=d$ case. This long cavity ($17.95\times 8.97\times 8975$ mm) can be operated in one of the two CAST magnet bores (43~mm in diameter and 9.3 m in length).

\begin{figure}[H]
	\includegraphics[width=0.99\textwidth]{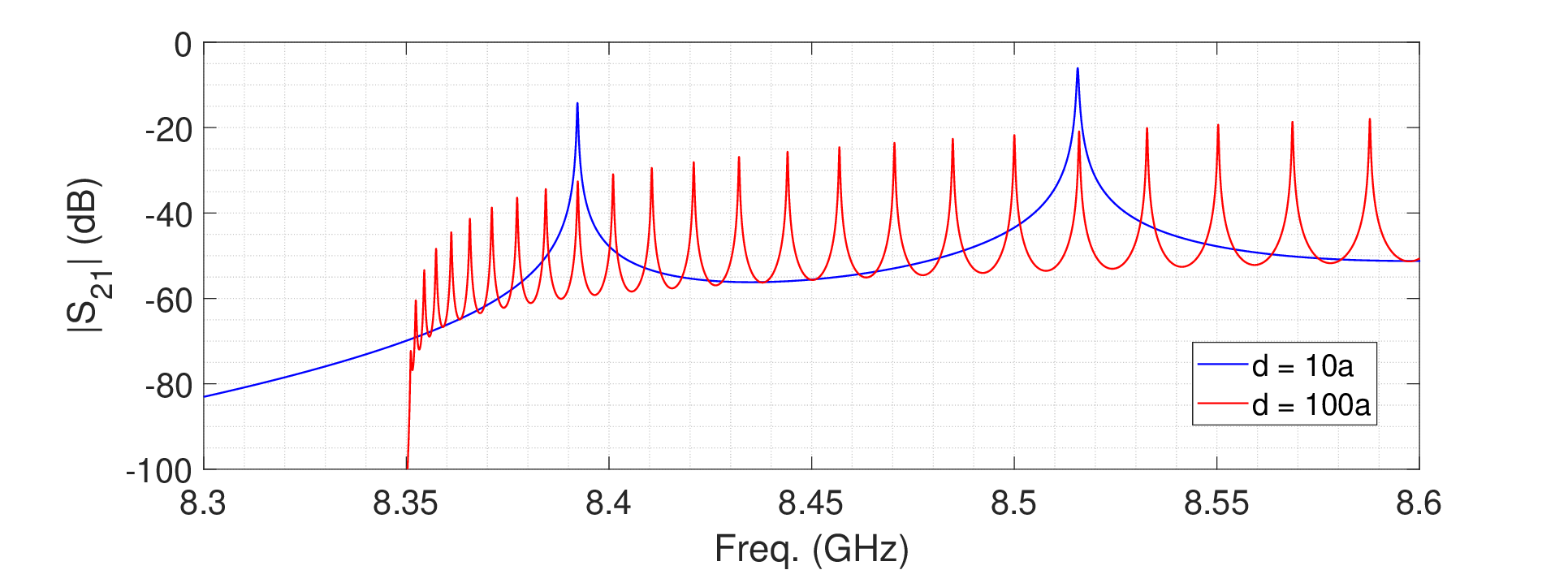}
    \caption{Cavity (17.95 × 8.97 cross-section) spectrum around 8.4 GHz for $d=10a$ (blue) and $d=100a$ (red). {Simulation results from CST Studio Suite \cite{CST}.}} 
	\label{fig:1cav_d10a&100a} 
\end{figure}

The great drawback of this length increment is the clustering of higher resonant modes $TE_{10p}$ close to the chosen axion-coupling mode ($TE_{101}$) \cite{Baker2018}. Figure \ref{fig:1cav_d10a&100a} shows the spectrum of the cavity around 8.4 GHz for different lengths. If the nearest higher mode ($TE_{102}$) is too close to $TE_{101}$, it can hinder the characterization of $Q_l$ of the constructed cavity, as well as the coupling factor $\kappa$.

\subsection{The Alternative: Multi-Cavity Concept}

Multi-cavity structures are widely used in filtering for telecommunication applications in the range of microwaves and millimeter waves. In that context, the aim is realizing a band pass filter with a given mask which prescribes the pass band, the stop band, the pass ripple, and the stop band attenuation. Different references \cite{Matthaei1980,Cameron2018} deal with the analysis and design of this kind of structure.

In our context, the objective is the design of a very selective filter (high $Q_l$), centered on the resonant frequency of interest, with a large volume and with a mode that couples optimally with the external magnetic field in order to maximize the form factor. The mathematical formalism for the analysis and design of this kind of multi-cavity device is presented in \cite{Alvarez2018}.

As explained there, roughly speaking, if the cavities have the same dimensions, the resonant frequency of the haloscope is that of the cell cavity, whilst the volume is the sum of volumes of the cell cavities. This means that, by designing properly a haloscope of $N$ cavities, the volume is multiplied by $N$ whilst the resonant frequency remains that of a single cavity. Moreover, as explained in \cite{Alvarez2018}, the quality factor of the haloscope must be approximately the same as that of the unit cavity, and the optimal external coupling factor $\kappa=\frac{1}{2}$ can be obtained by extracting the signal from any of the cavities.

Assuming that other modes are sufficiently far from the $TE_{101}$ mode, all cavities will resonate in that mode, but the whole device can have different mode configurations, depending on the couplings between neighboring cavities. The chosen mode configuration will be that which maximizes the form factor, which will be the configuration with all positive (or all negative) electric fields in the cavities. This allows a constructive sum in the term $|\int \vec{E}\cdot \vec{B}_edv|^2$ of the factor form (\ref{C}). In fact, it is expected that the total form factor will be approximately the same as one of the unit cavities, which for the $TE_{101}$ can be obtained analytically from (\ref{C}) as
\begin{equation}
    C_{TE_{101}} = \frac{|\int\sin{\left(\frac{\pi x}{a}\right)}\sin{\left(\frac{\pi z}{d}\right)}dxdydz|^2}{V\int\sin^2{\left(\frac{\pi x}{a}\right)}\sin^2{\left(\frac{\pi z}{d}\right)}dxdydz}=\frac{64}{\pi^4}
\end{equation}

This multi-cavity concept is very close to the idea of coherently summing $N$ signals extracted from $N$ identical cavities, but with only one compact device and avoiding the difficulty of conserving the coherence (in-phase signals) in the combining device. Other multi-cavity concepts for axion detection have been published in the last few years \cite{Goryachev2018,Jeong2020}.

\vspace{-6pt}

\begin{figure}[ht!] 
	        \includegraphics[width= 0.49\textwidth]{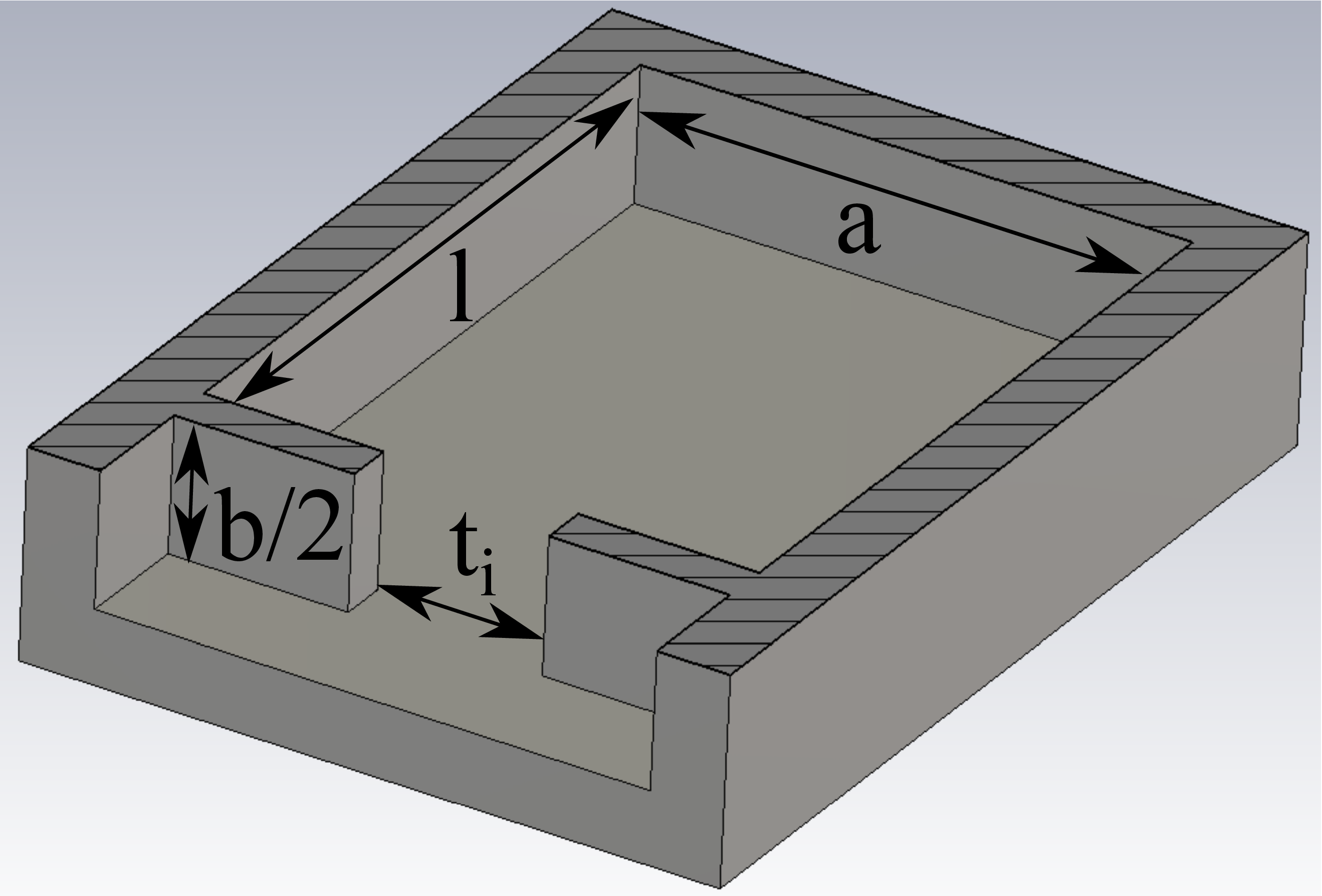}
	        \includegraphics[width=0.49\textwidth]{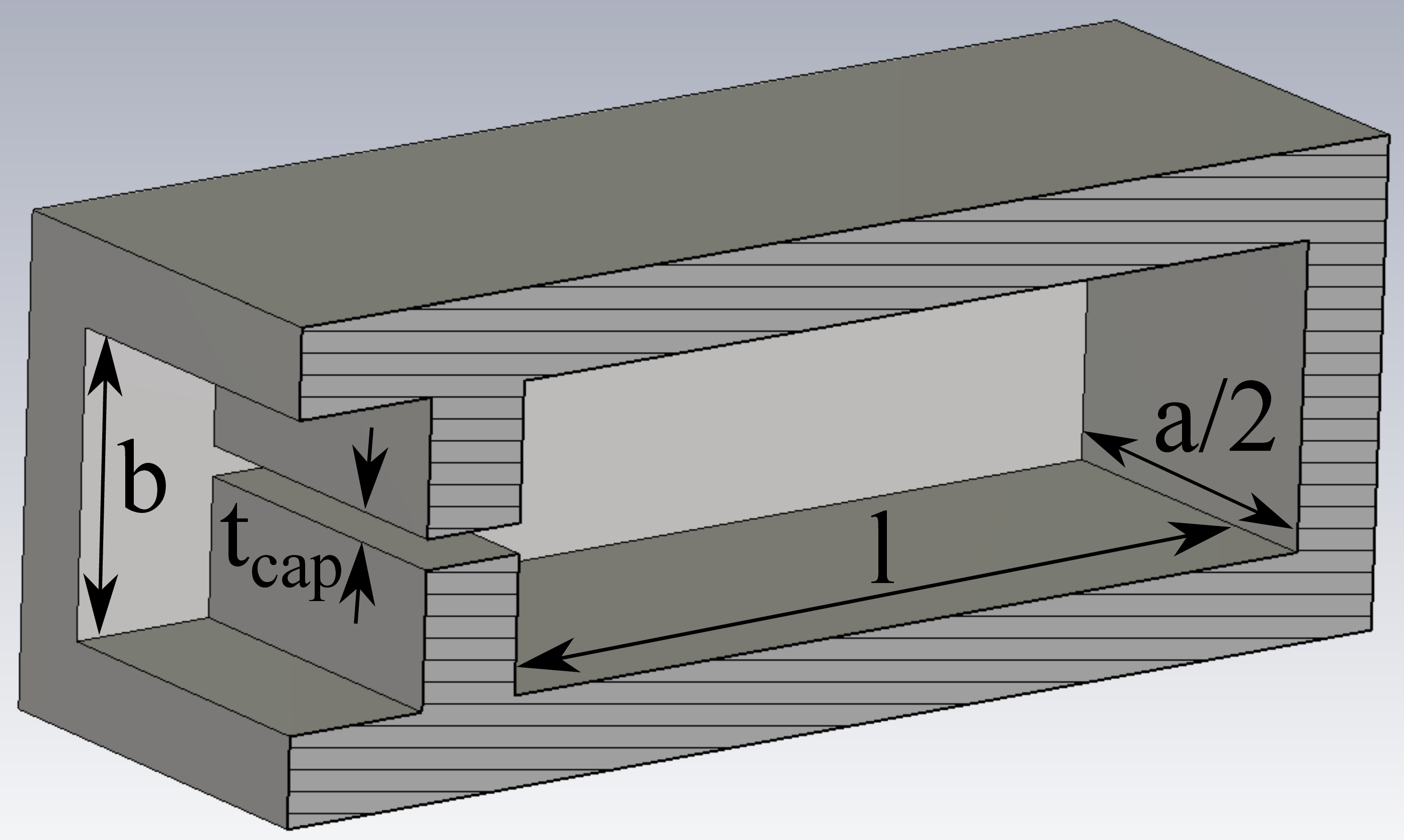}
		\caption{Rectangular irises for coupling between neighboring cavities: inductive (\textbf{left}) and capacitive (\textbf{right}). The pictures show the symmetric half of each one.}
		\label{fig:Ind&Cap}
\end{figure}

In RADES, the coherent construction of the all-positive electric field configuration is obtained through the proper design of the internal couplings. These couplings between neighboring cavities can be realized, in the simplest way, with rectangular irises, such as those shown in Figure \ref{fig:Ind&Cap}. Due to the vertical polarization of the $TE_{101}$ electric field, irises invariant in $y$ (Figure \ref{fig:Ind&Cap} left) have an inductive behavior, whilst those invariant in $x$ (Figure \ref{fig:Ind&Cap} right) have a capacitive behavior.

\section{First Multi-Cavity Haloscopes}
\subsection{All Inductive 5-Cavity Haloscope}
\label{subsection:All inductive 5-cavities haloscope}

\begin{figure}[ht!]
	        \includegraphics[width= 0.49\textwidth]{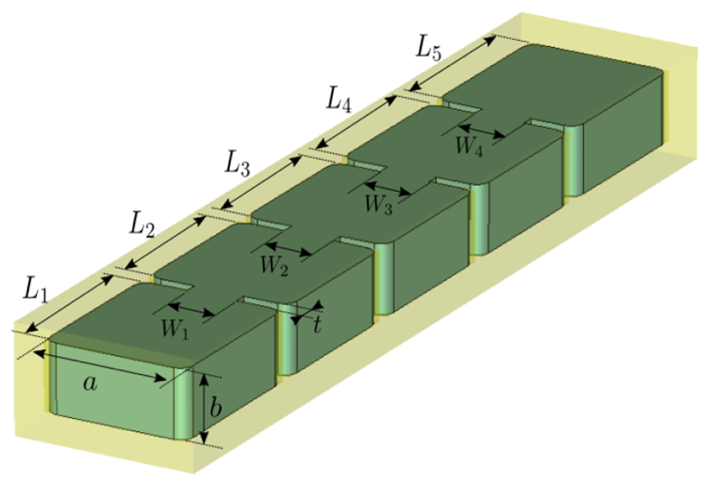}
	        \includegraphics[width= 0.49\textwidth]{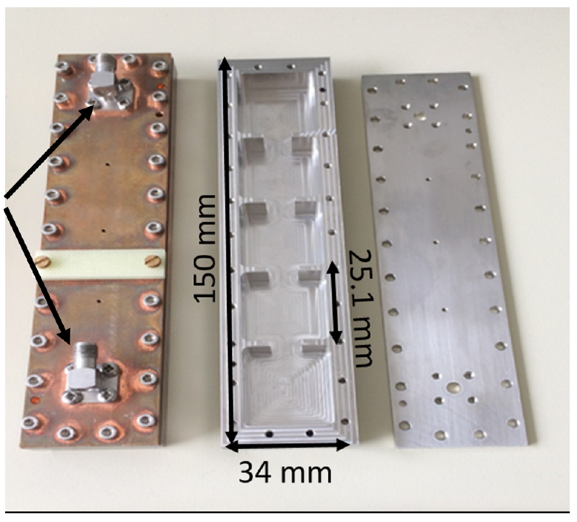}\label{fig:5cav_a}
		\caption{RADES 5-cavity haloscope: scheme and dimensions, and device before copper plating.}
		\label{fig:5cav}
\end{figure}

During 2017, the first RADES haloscope was designed, manufactured, characterized, and finally used in the 2018 data-taking campaign in CAST. It was the fivefold cavity shown in Figure \ref{fig:5cav} and presented in \cite{Alvarez2018}. Dimensions, as described in Figure \ref{fig:5cav}, are listed in Table \ref{table:physdimen5cav}.

\begin{figure}[H]
	\includegraphics[width=0.9\textwidth]{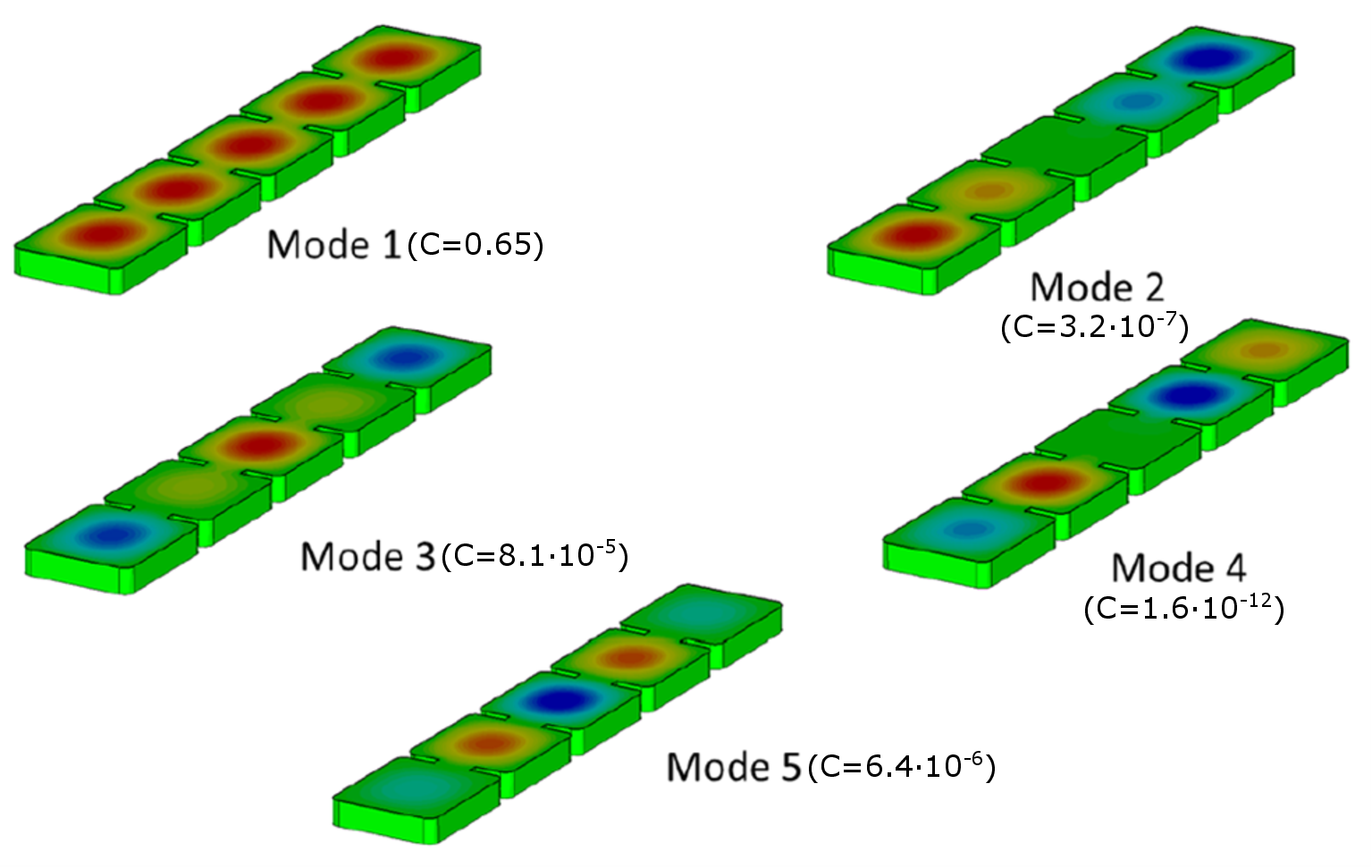}
    \caption{Electric field pattern (vertical polarization) for the five different configurations of mode $TE_{101}$ for the 5-cavity haloscope, with the associated form factors. The red fields denote positive levels, the green fields zero, and the blue fields negative. Taken and modified from \cite{Alvarez2018}.}
	\label{fig:5cav_Modes} 
\end{figure}

This device was designed first through the formalism in \cite{Alvarez2018}, and was finally optimized with CST Studio \cite{CST}. Figure \ref{fig:5cav_Modes} shows the five different configurations of the mode $TE_{101}$ for the final five-cavity structure, and the corresponding form factors.

\begin{figure}[H]
	\includegraphics[width=0.9\textwidth]{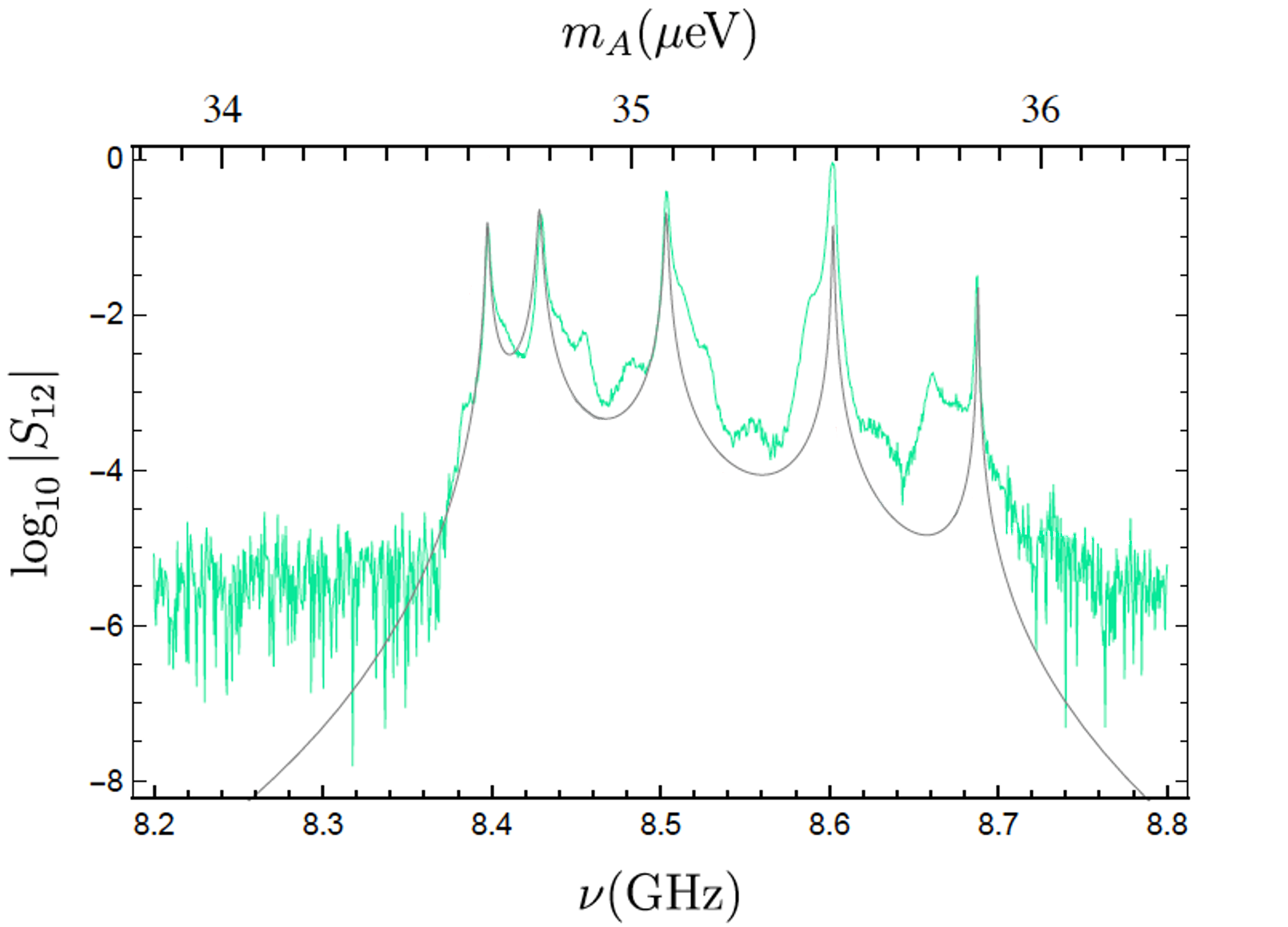}
    \caption{Transmission coefficient magnitude at 2 K: simulated (black) and measured (green), where measurements include the effects of cables from and to the VNA. Taken and modified from \cite{Alvarez2018}.}
	\label{fig:5cav_VNA_simulated}
\end{figure}

Occasional quenches produced in the CAST magnet made necessary a strong material for the haloscope in order to avoid deformations in the device. For that reason, the base material was not copper, but non-magnetic stainless steel 316LN, plated with a 30 $\upmu$m of copper layer for reducing the losses. The transmission coefficient magnitude, both simulated and measured at the operation temperature of 2 K, is depicted in Figure \ref{fig:5cav_VNA_simulated}.

\vspace{-9pt}
\begin{table}[H]
	\caption{\label{table:physdimen5cav} Physical dimensions of the 5-cavity haloscope at 2 K and at 298 K.} 
	\setlength{\tabcolsep}{7.8mm}
	\begin{tabular}{ccc}
		\toprule
		\textbf{Dimensions (mm)}  & \boldmath{$T = 2$} \textbf{K}  & \boldmath{$T = 298$} \textbf{K}   \\ \midrule
		Cavity width ($a$) & 22.86 & 22.99 \\ 
		Cavity height ($b$) & 10.16 & 10.25 \\ 
		Length external cavities ($L_1=L_5$) & 26.68 & 26.82 \\ 
		Length internal cavities ($L_2=L_3=L_4$) & 25.00 & 25.14 \\ 
		Iris width ($W_1=W_2=W_3=W_4$) & 8.00 & 8.14 \\ 
		Iris thickness ($t$) & 2.00 & 1.95 \\ \bottomrule
	\end{tabular}
\end{table}

\vspace{-16pt}

\vspace{-8pt}

\subsection{Alternate Coupling}

\begin{figure}[H]
	\includegraphics[width=0.9\textwidth,height=300pt]{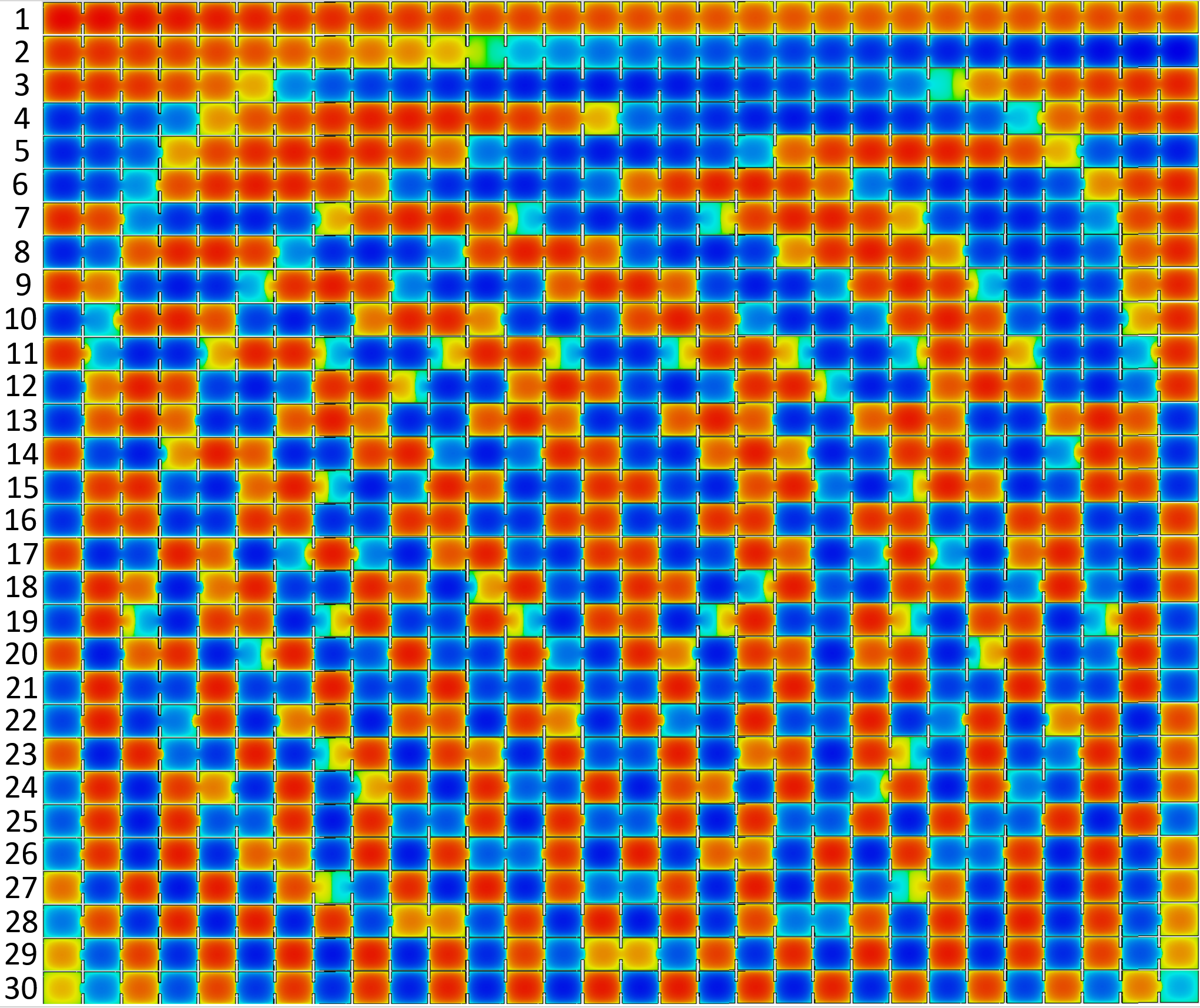}
    \caption{Electric field pattern (vertical polarization) for the thirty different configurations of mode $TE_{101}$ for the all-inductive 30-cavites haloscope. Numbers on left side refer to the order of the configuration resonances with the frequency. {The red regions denote positive E-fields, and the blue regions negative ones.}}
	\label{fig:30cavAllind_Modes_E-fieldPattern}
\end{figure}

\begin{figure}[H]
	\includegraphics[width=0.9\textwidth]{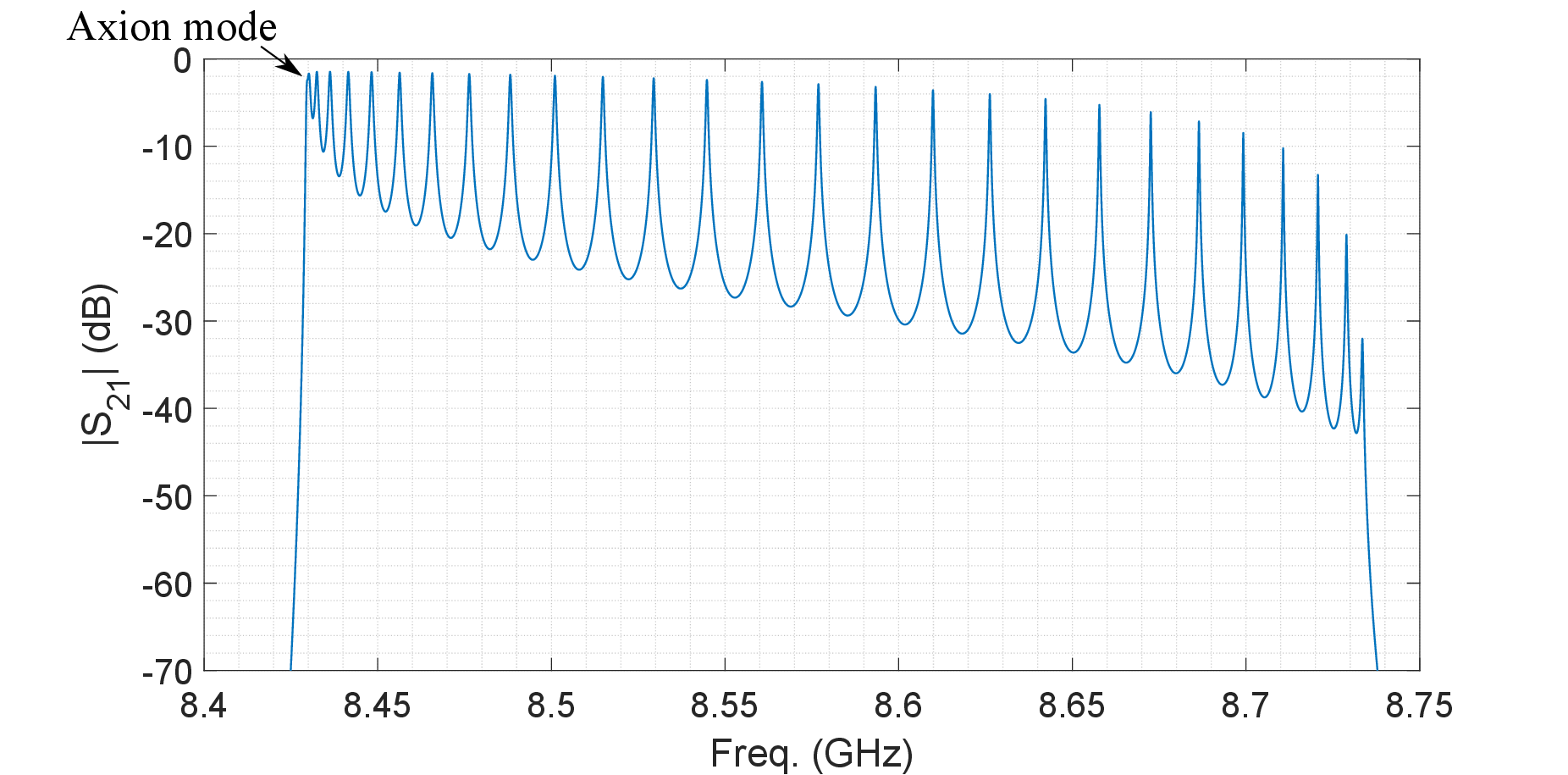}
    \caption{\textls[-15]{Simulated transmission coefficient magnitude of an all-inductive 30-cavity haloscope at 2 K.}}
	\label{fig:30cav_allind_S21_Cu2K} 
\end{figure}

Although the multicavity concept alleviates the mode clustering, it has also a limit. N cavities produce N different configuration patterns for $TE_{101}$ mode. Therefore, when the number of cavities increases, the same concentration of resonances is observed. \mbox{Figure \ref{fig:30cavAllind_Modes_E-fieldPattern}} shows the 30 configurations of a 30-cavity haloscope with a similar design to the 5-cavity one, i.e., inductive irises. Figure \ref{fig:30cav_allind_S21_Cu2K} shows the spectrum of this device with \mbox{these resonances.}

\vspace{-8pt}

As observed, the first (axion one) and the second $TE_{101}$ mode configurations are very similar. This situation worsens with the manufactured device and data-taking conditions, leading normally to an overlapping of both resonances, which makes it impossible to obtain the quality factor or the coupling factor from measurements.

\begin{figure}[H]
	\includegraphics[width=0.9\textwidth,height=300pt]{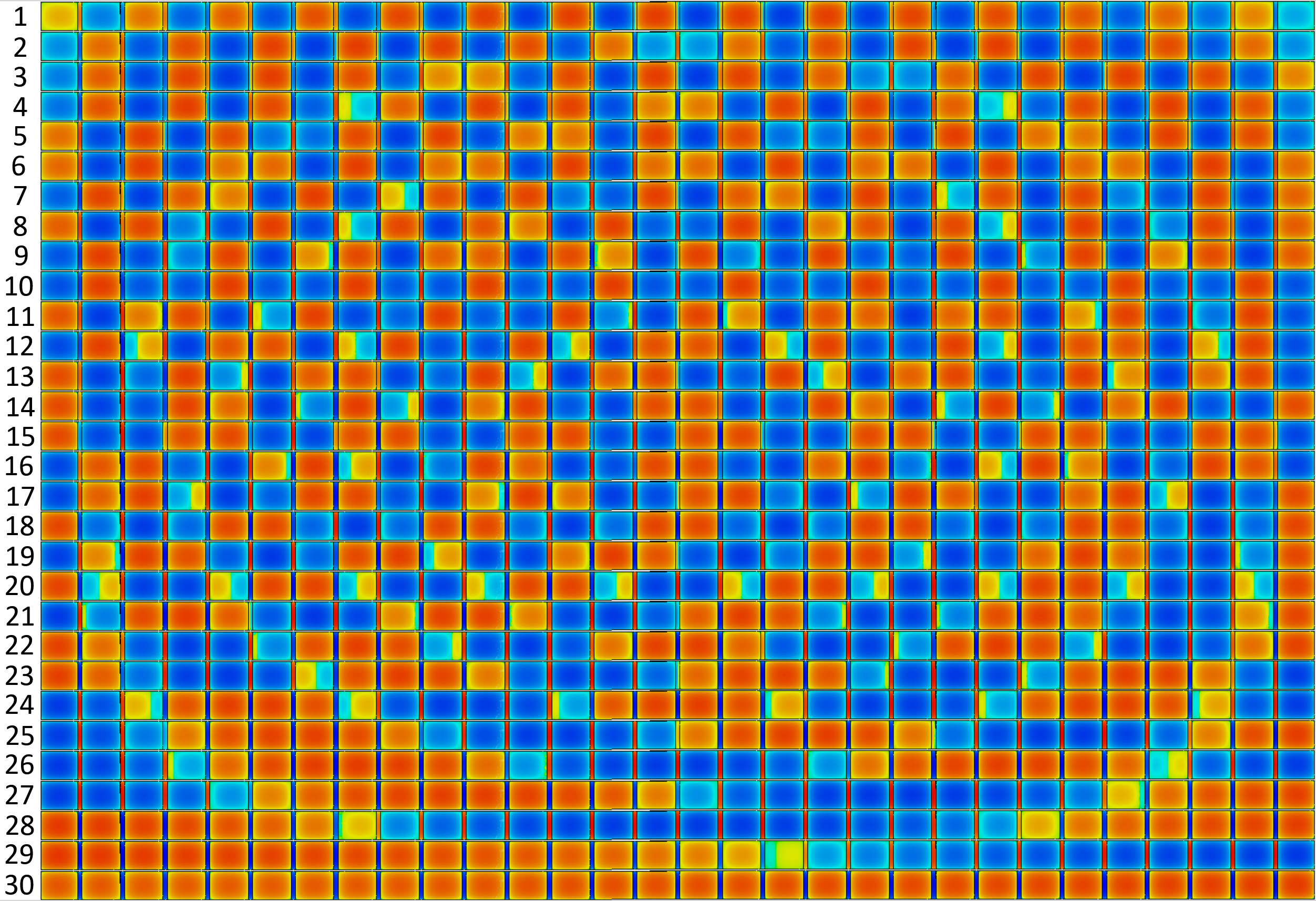}
    \caption{Electric field pattern (vertical polarization) for the thirty different configurations of mode $TE_{101}$ along the all-capacitive 30-cavites haloscope. Numbers on the  left side refer to the order of the configuration resonances with the frequency. {The red regions denote positive E-fields, and the blue regions negative ones.}}
	\label{fig:30cavAllcap_Modes_E-fieldPattern}
\end{figure}

\begin{figure}[H]
	\includegraphics[width=0.9\textwidth]{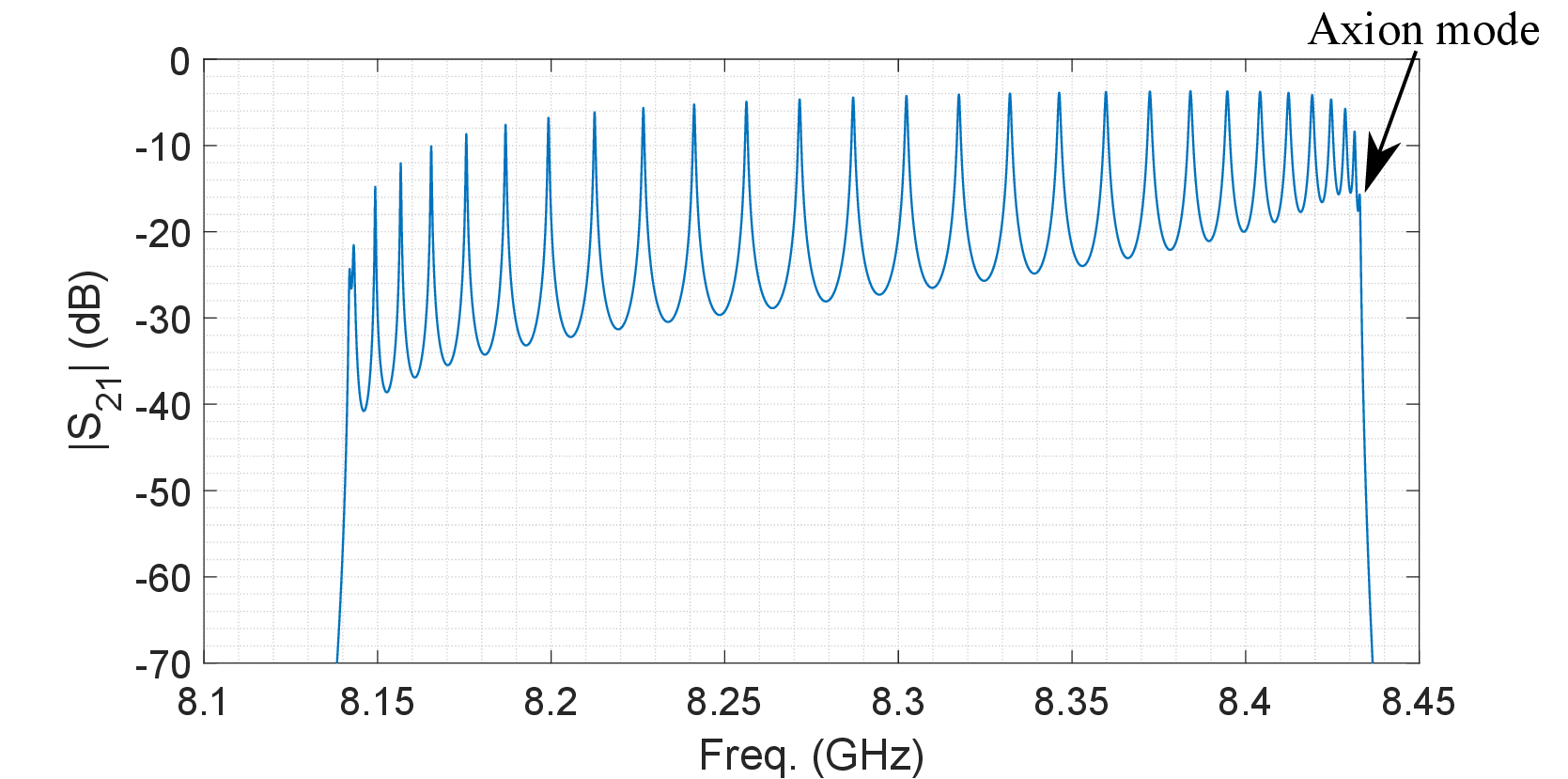}
    \caption{\textls[-25]{Simulated transmission coefficient magnitude of an all-capacitive 30-cavity haloscope at 2 K.}} 
	\label{fig:30cav_allcaps_S21_Cu2K}
\end{figure}

\textls[-15]{For an all-capacitive coupling, the patterns and spectrum are shown in \mbox{Figures \ref{fig:30cavAllcap_Modes_E-fieldPattern} and \ref{fig:30cav_allcaps_S21_Cu2K}}}, respectively. In both spectrums, thirty modes can be seen (not by eye due to numerical accuracy from the software). However, there is a main difference between them: for the all-inductive structure, the axion mode is the first peak, whereas for the all-capacitive structure, the axion mode is the last one (30th mode). Note that in both cases, the axion mode is at the same frequency ($\sim$8.43 GHz). The reason comes from the design parameters employed to set such an axion search at that specific frequency.

\vspace{-15pt}

\begin{figure}[H]
	\includegraphics[width=0.9\textwidth]{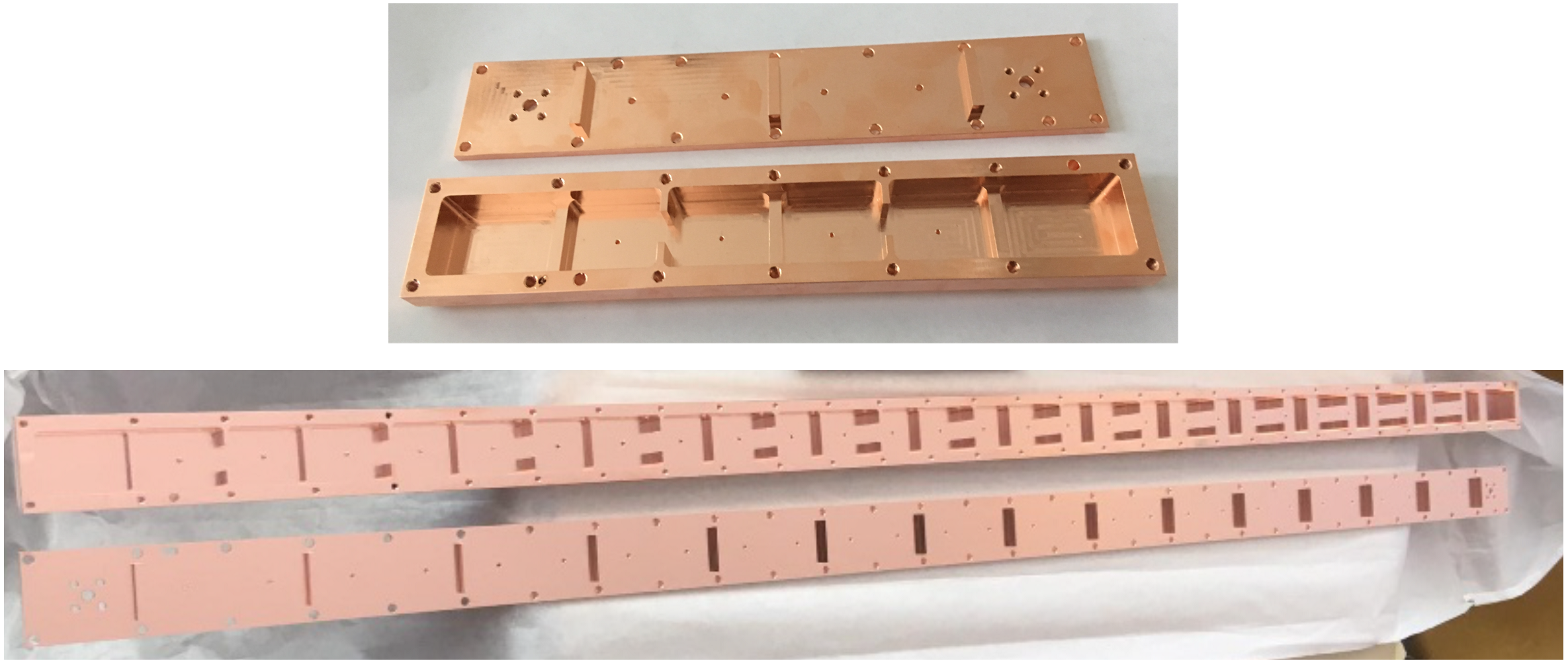}
    \caption{Manufactured 6cav and 30cav structures with alternated couplings.}
	\label{fig:6-and-30CAV-ALT-FABRICADA}
\end{figure}

\begin{figure}[H]
	\includegraphics[width=0.9\textwidth,height=300pt]{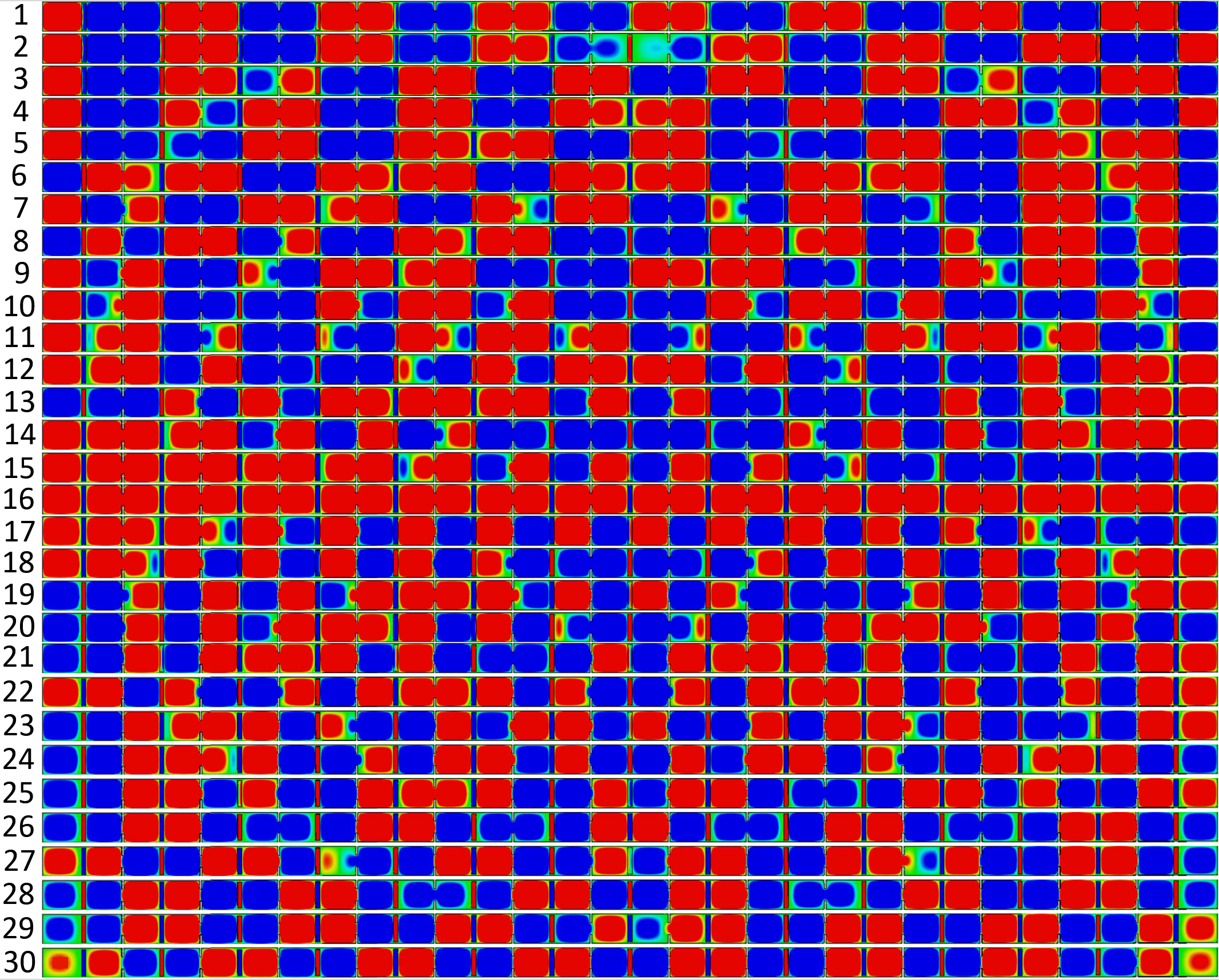}
    \caption{Electric field pattern (vertical polarization) for the thirty different configurations of mode $TE_{101}$ for the alternating 30-cavites haloscope. Numbers on the left refer to the order of the configuration resonances with the frequency. {The red regions denote positive E-fields, and the blue regions negative ones.}}
	\label{fig:30cav_alt_Modes_E-fieldPattern}
\end{figure}

\begin{figure}[H]
	\includegraphics[width=0.9\textwidth]{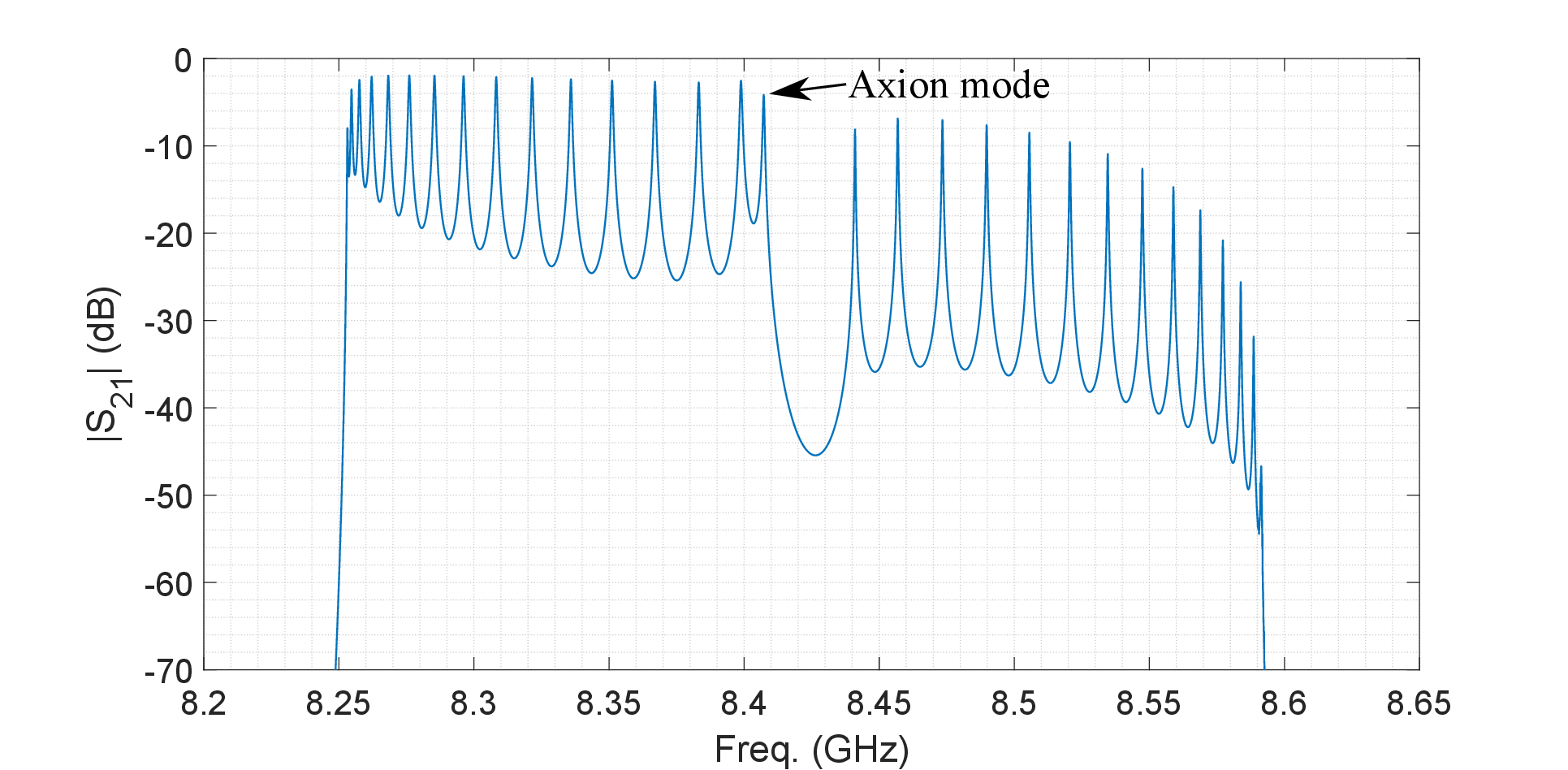}
    \caption{\textls[-15]{Simulated transmission coefficient magnitude of an alternating 30-cavity haloscope at 2 K.}}
	\label{fig:30cav_alt_S21_Cu2K} 
\end{figure}

The solution to avoid the mode-mixing close to the axion mode is to apply alternated couplings: capacitive + inductive irises \cite{Alvarez2020}. In this configuration the axion mode is at the middle of the band, where greater mode separation exists. The first haloscope designed and manufactured with this alternating concept was a 6-cavity structure (Figure \ref{fig:6-and-30CAV-ALT-FABRICADA} (top)), followed by a 30-cavity structure based on the same idea (Figure \ref{fig:6-and-30CAV-ALT-FABRICADA} (bottom)). Figure \ref{fig:30cav_alt_Modes_E-fieldPattern} shows the thirty E-field patterns of this last haloscope,  the sixteenth configuration being the most adequate for the axion--photon coupling, that is, with maximum form factor. The spectrum of this structure in the frequency range of interest is shown in Figure \ref{fig:30cav_alt_S21_Cu2K}, where an improvement regarding the mode-mixing can be observed. Concretely, the alternate coupling allows a 32-fold greater separation between the axion mode and the closest neighbor (from $\Delta f_{all-inductive} = 470$ KHz to $\Delta f_{alternated} = 15.34$ MHz).

\vspace{-9pt}

\section{Data Acquisition System}
\label{section:Data Acquisition system}

\begin{figure}[H]
 \includegraphics[height= 8cm,width = 0.9\textwidth]{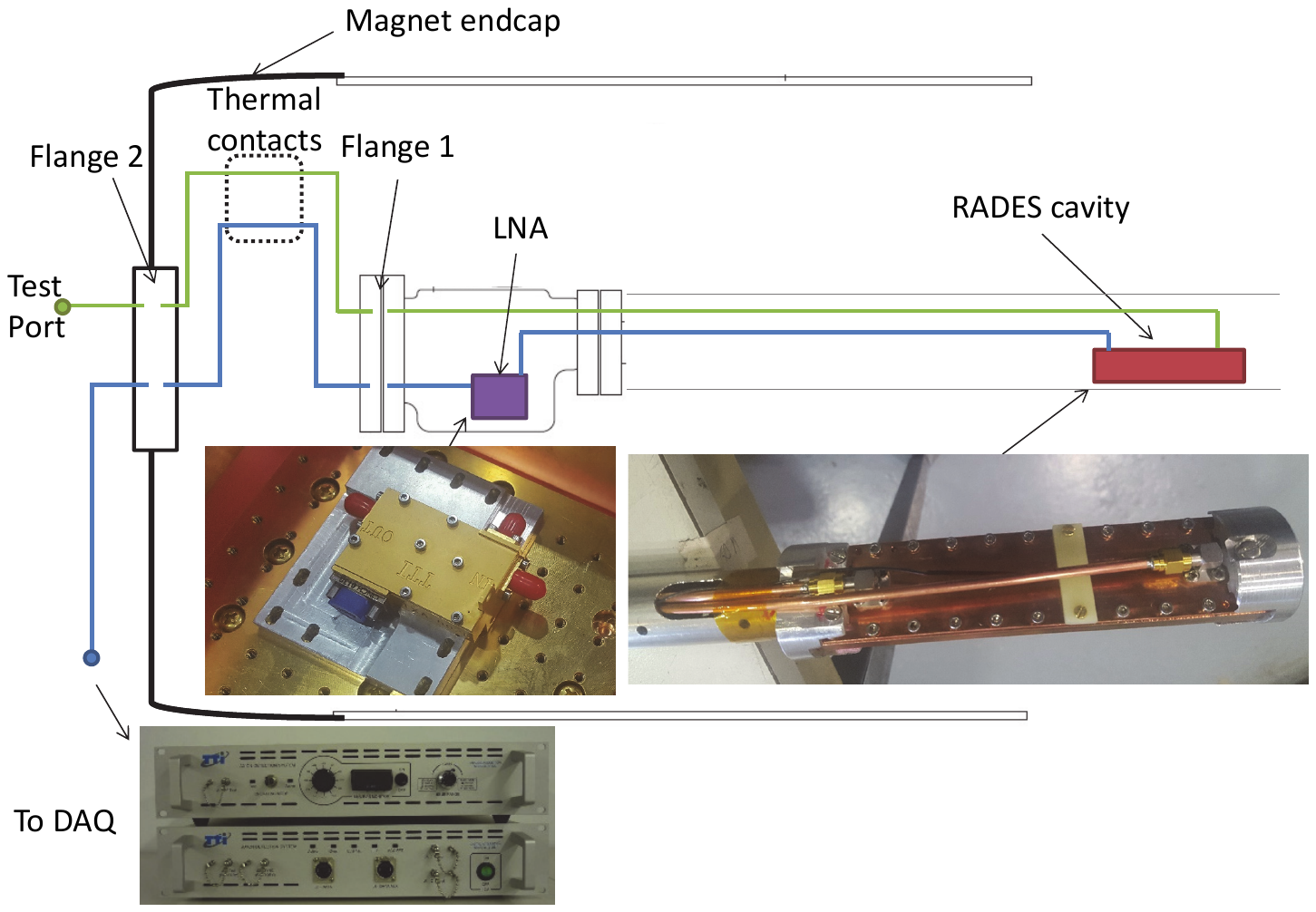}
 \caption{Measurement setup. The cavity was placed in the cold bore at near 2 K. The signal was amplified by a cryogenic LNA. Thermal contacts were employed to adapt the cryogenic and room temperatures for both the calibration and signal ports between flanges 1 and 2.  }\label{wsetup}
\end{figure}

The setup for the experiment is shown in Figure \ref{wsetup}. The cavity was placed inside one of the bores of the Large Hadron Collider (LHC) ring section prototype placed at CAST. A low noise amplifier (LNA) TXA4000 from TTI Norte \cite{tti} provided a 40 dB gain in the 8--9 GHz frequency band and was placed inside a copper vessel in the cryogenic section limited by flange 1. Then a thermal transition required thermal plates to adapt the cables from the cryogenic section to room temperature. Port 1 of the haloscope was designed to obtain critical coupling while port 2, which was weakly coupled, provided information at the calibration stage about the working frequency and the correct operation of the amplifier, which needed a special bias. Temperature and bias cables were made of phosphor bronze from Lake Shore Cryotronics \cite{lakeshore} to avoid thermal leakages. Radiofrequency (RF) cables were 3.5 mm semirigid coaxial copper from Micro-coax \cite{microcoax}. Connectors were SubMiniature version A (SMA).

\vspace{-4pt}

\begin{figure}[H]
 \includegraphics[height= 4cm,width = 0.9\textwidth]{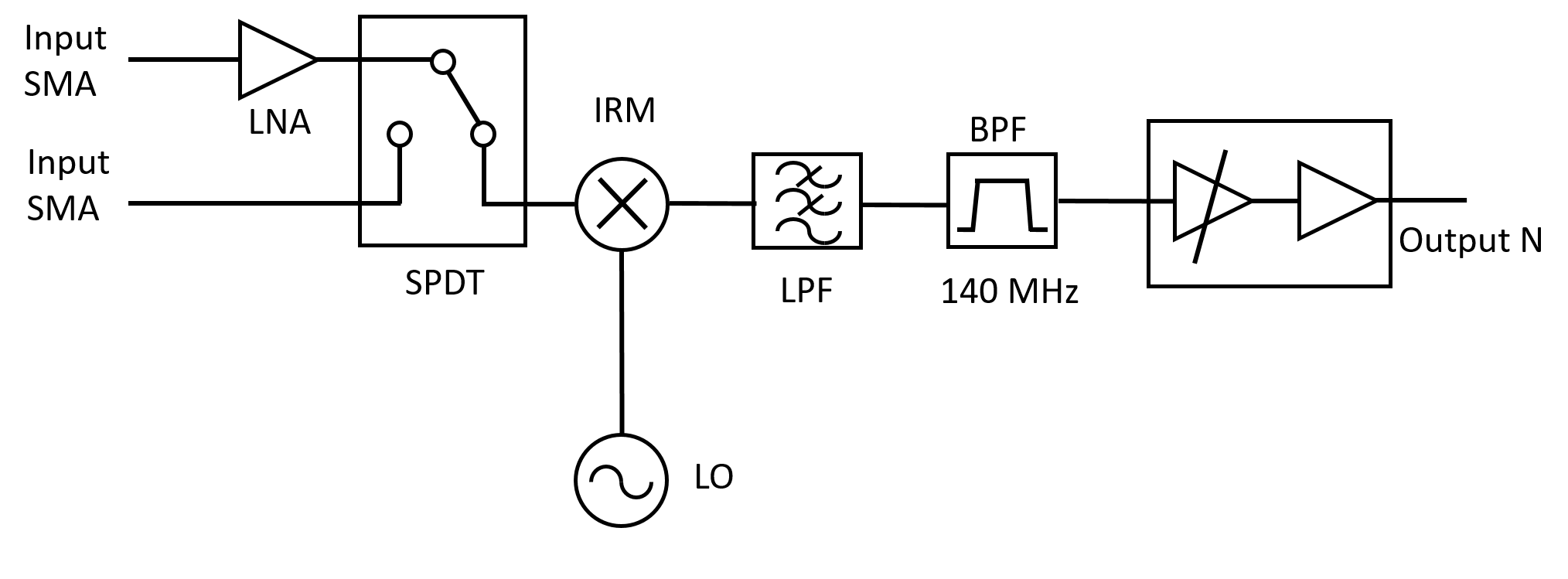}
 \caption{Block diagram of the analog acquisition module.}\label{daqa}
\end{figure}

The signal from port 1 was the input of the analog acquisition module, which was a heterodyne receiver that down-converted the input RF signal to the intermediate frequency (IF) output frequency band. The main functional blocks of the analog acquisition module were (see Figure \ref{daqa}):

\begin{itemize}
	\item RF LNA: It amplified the input RF signal. It worked from $8$ GHz to $9$ GHz and provided a gain around $55$ dB and $30$ dB input return loss.
	
	\item Single pole double through (SPDT) RF switch: This device allowed us to visualize the amplified RF signal at the test output port, or to  directly check the down \mbox{converter circuit.} 
				
	\item Down converter: It down-converted the input signal from X band to a frequency band around $140$ MHz. It was an image rejection mixer (IRM) which provided $26$ dB of image band rejection.
	
	\item Local oscillator (LO): It provided the frequency signal to convert the RF input signal to the $140$ MHz band. Its working power level was $0$ dBm working from $7860$  to $8860$ MHz.
	
	\item IF filtering: This section filtered the desired IF signal from $134$ MHz to $146$ MHz. It consisted of two filters in series: a low pass filter (LPF) to reject the LO leakage from the IRM, and a surface acoustic wave (SAW) band pass filter (BPF) to select the wanted IF frequency band.
	
	\item IF signal conditioning: This module amplified or attenuated the IF signal to provide the desired IF output level. Its function was to bring the IF signal to the best input level for the digital acquisition module. The amplification or attenuation level depended on the input RF signal power.
	
	\item Power supply unit: This unit fed all the modules that formed the analog \mbox{acquisition module.}
\end{itemize}

The LO frequency was chosen to satisfy (\ref{IF}), taking into account the radio-frequency (frequency of the first appearing mode of the cavity).
\begin{equation} \label{IF}
    \text{IF(MHz)} \, = \, \text{RF(MHz)} \, - \, \text{LO(MHz)}
\end{equation}

\noindent(for instance, $140$ MHz $=$ $8400$ MHz $-$ $8260$ MHz) 

\begin{figure}[H]
 \includegraphics[height= 3cm,width = 0.9\textwidth]{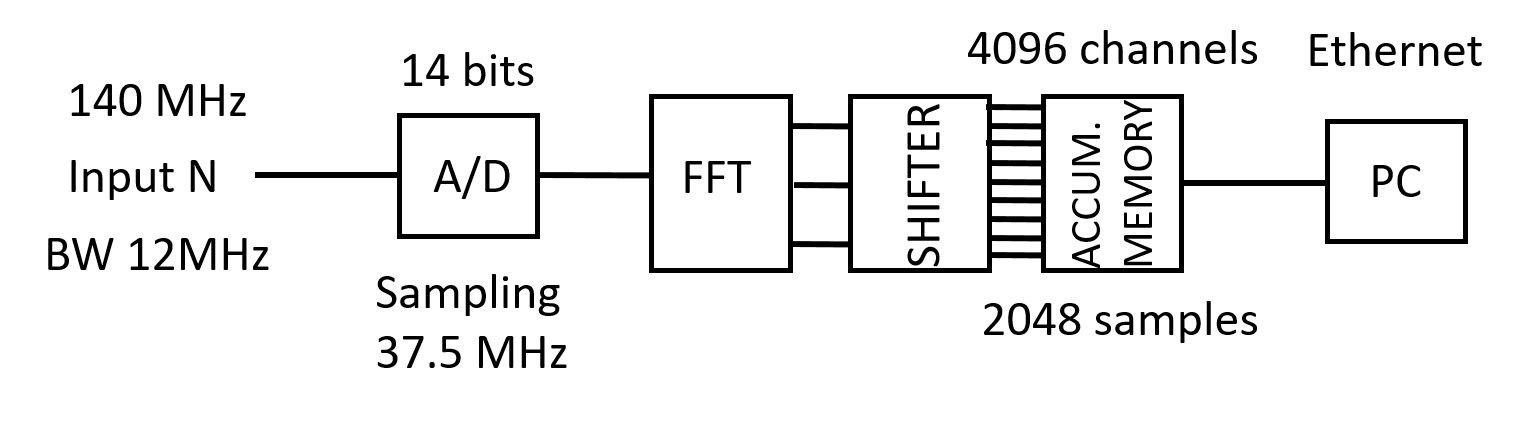}
 \caption{Block diagram of the digital acquisition module.}\label{daqd}
\end{figure}

The digital acquisition module represented in Figure \ref{daqd} was especially designed to process the IF signal to obtain the power spectra with an adequate dynamic range and frequency resolution. This is achieved by digitizing the signal with a $14$ bit A/D converter and processing the data with a FPGA. The unit analyzes a total instantaneous bandwidth of $12$ MHz around a central frequency of $140$ MHz. This is performed by sampling with a clock frequency of $37.5$ MHz and using an antialiasing filter centered in the eighth Nyquist zone (bandpass sampling). Then, the FFT is calculated with $8192$ samples, obtaining a frequency resolution of $4577.6$ Hz. The system is designed to integrate (average) a number of the power spectra internally to avoid unnecessary large data rates and file storage requirements in the external computer. A pre-integrated spectrum is sent every $0.44739$ seconds to be further integrated by the data acquisition software.\\

\section{First Results}

In 2018 the 5-cavity haloscope described in Section \ref{subsection:All inductive 5-cavities haloscope} and the associated receiver were installed in the CAST magnet. The data-taking period went from the 30th of October of 2018 to the 20th of December of 2018. During this time, 432 h of magnet-on and 141 h of magnet-off data at different LO frequencies were taken. The reason for taking different LO frequencies will be explained later on in this section.

A typical power spectrum consisted of data integrated for 90.37278 s and is labeled a 90s spectrum. For the first results, two sets of magnet-on and two sets of magnet-off data were used. The magnet-on sets were taken at LO frequencies, $l_1$ = 8.240 GHz and $l_{2-\text{on}}$ = 8.247 GHz, and contained 4093 and 2446 spectra, respectively. The magnet-off sets were taken at LO frequencies $l_1$ and $l_{2-\text{off}}$ = 8.248 GHz, and contained 1047 and 702 spectra, respectively. 

\begin{figure}[!ht]
\begin{subfigure}{.49\textwidth}
  \centering
  \includegraphics[width=.8\linewidth]{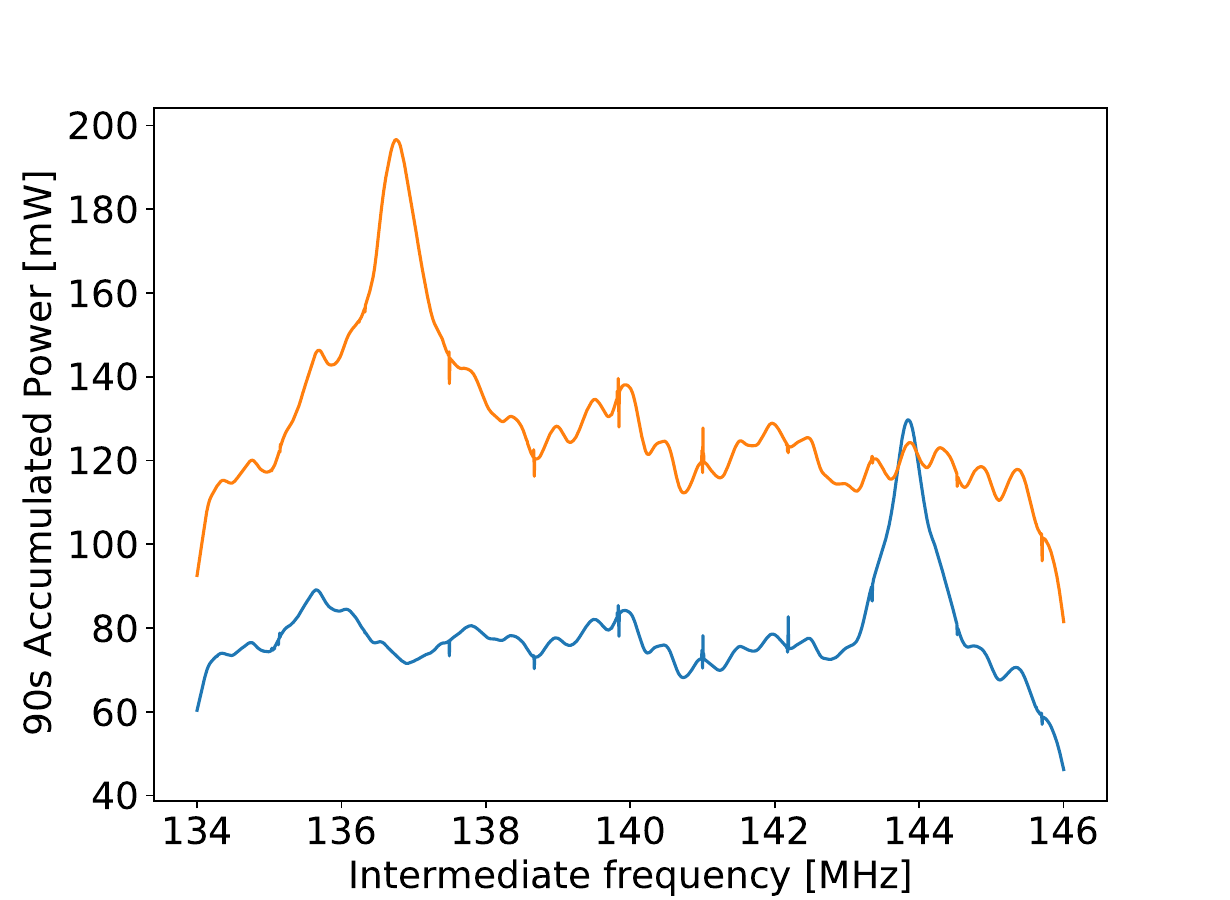}
  \caption{}
  \label{fig:Background-Removal-a}
\end{subfigure}%
\begin{subfigure}{.49\textwidth}
  \centering
  \includegraphics[width=.8\linewidth]{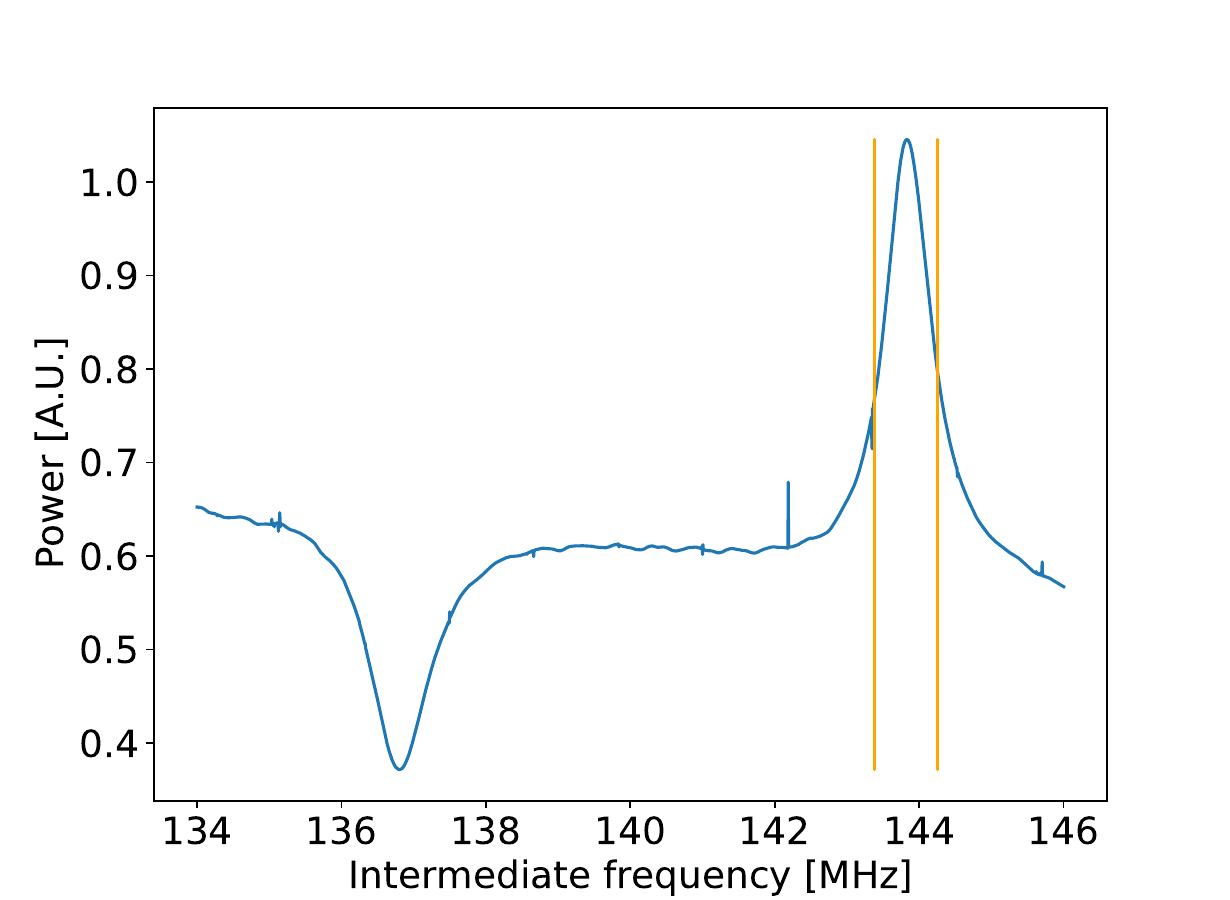}
  \caption{}
  \label{fig:Background-Removal-b}
\end{subfigure}
\begin{subfigure}{.49\textwidth}
  \centering
  \includegraphics[width=.8\linewidth]{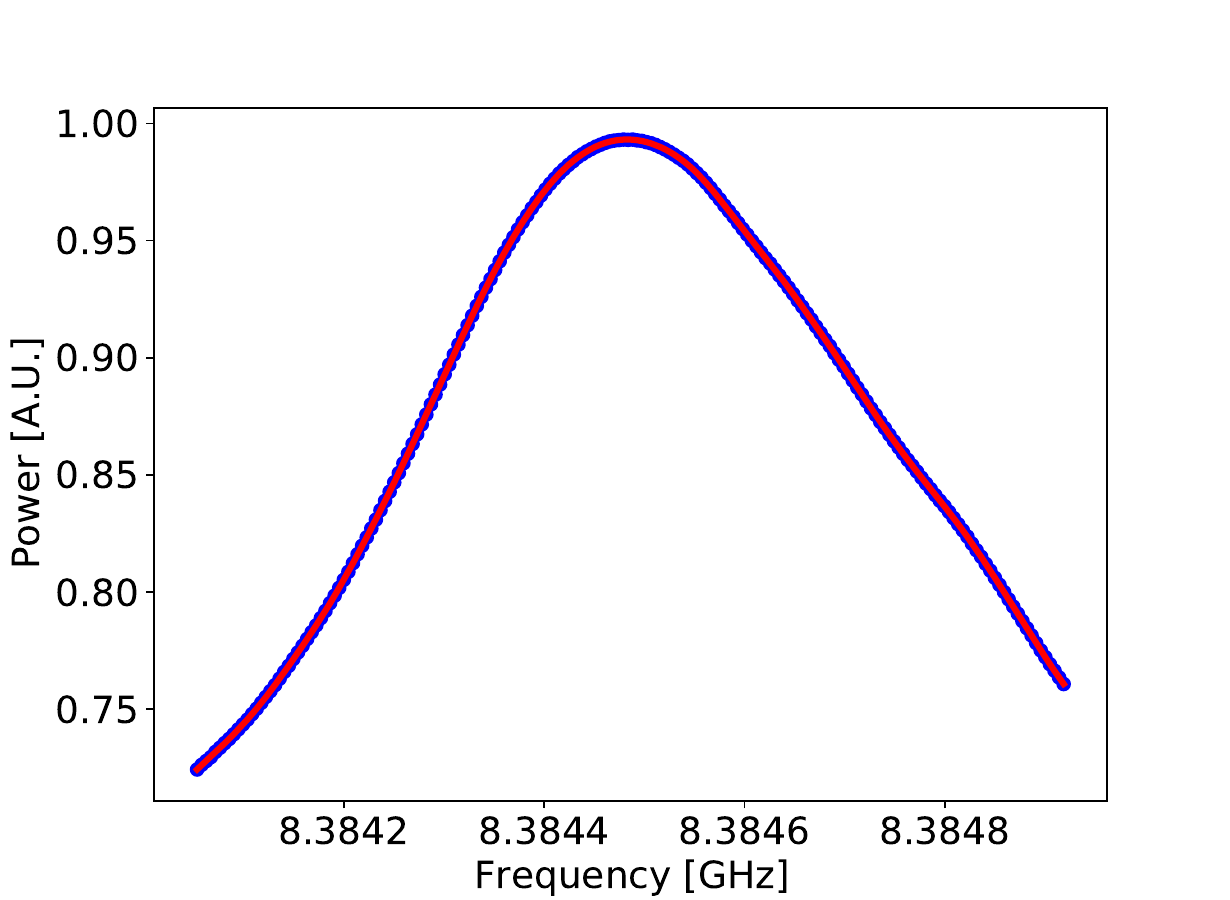}
  \caption{}
  \label{fig:Background-Removal-c}
\end{subfigure}%
\begin{subfigure}{.49\textwidth}
  \centering
  \includegraphics[width=.8\linewidth]{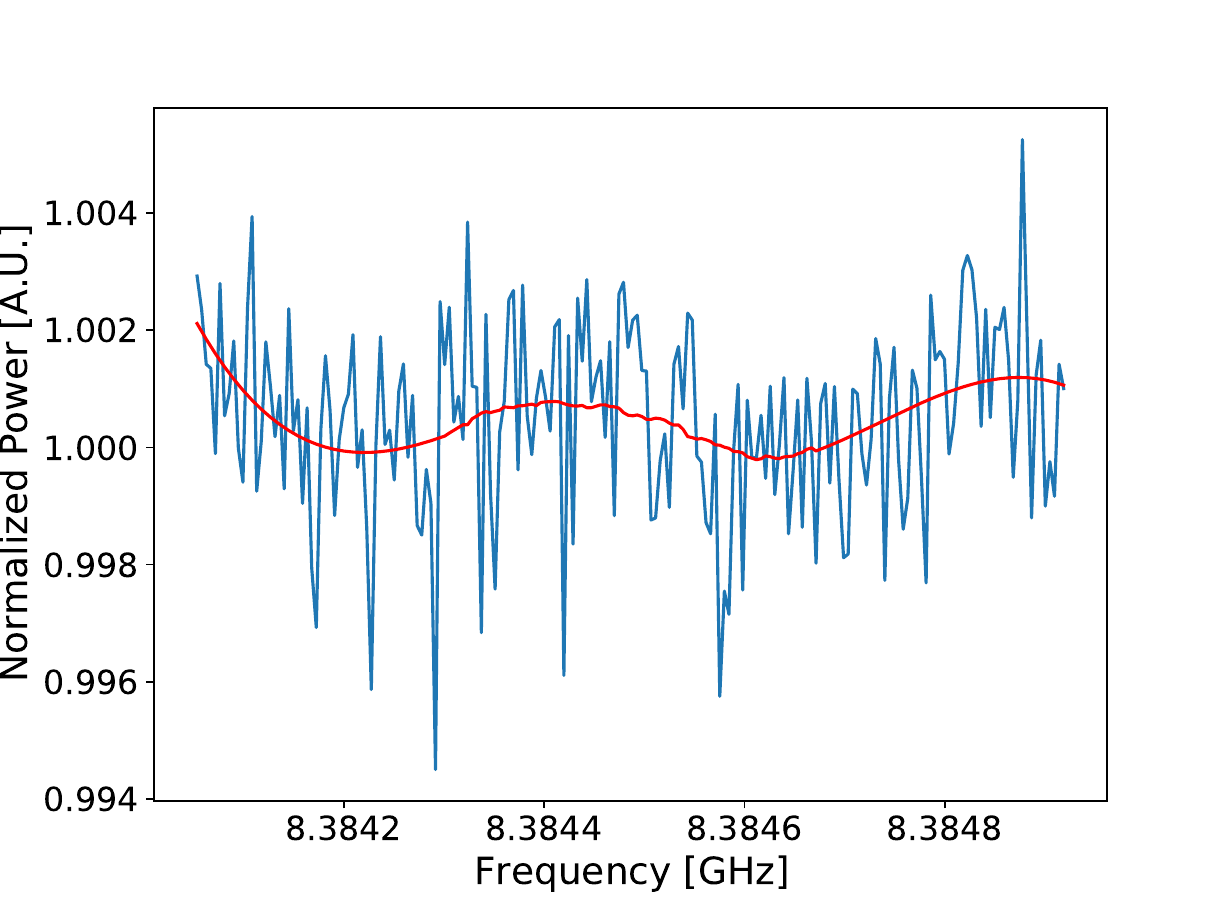}
  \caption{}
  \label{fig:Background-Removal-d}
\end{subfigure}

\begin{subfigure}{.49\textwidth}
  \centering
  \includegraphics[width=.8\linewidth]{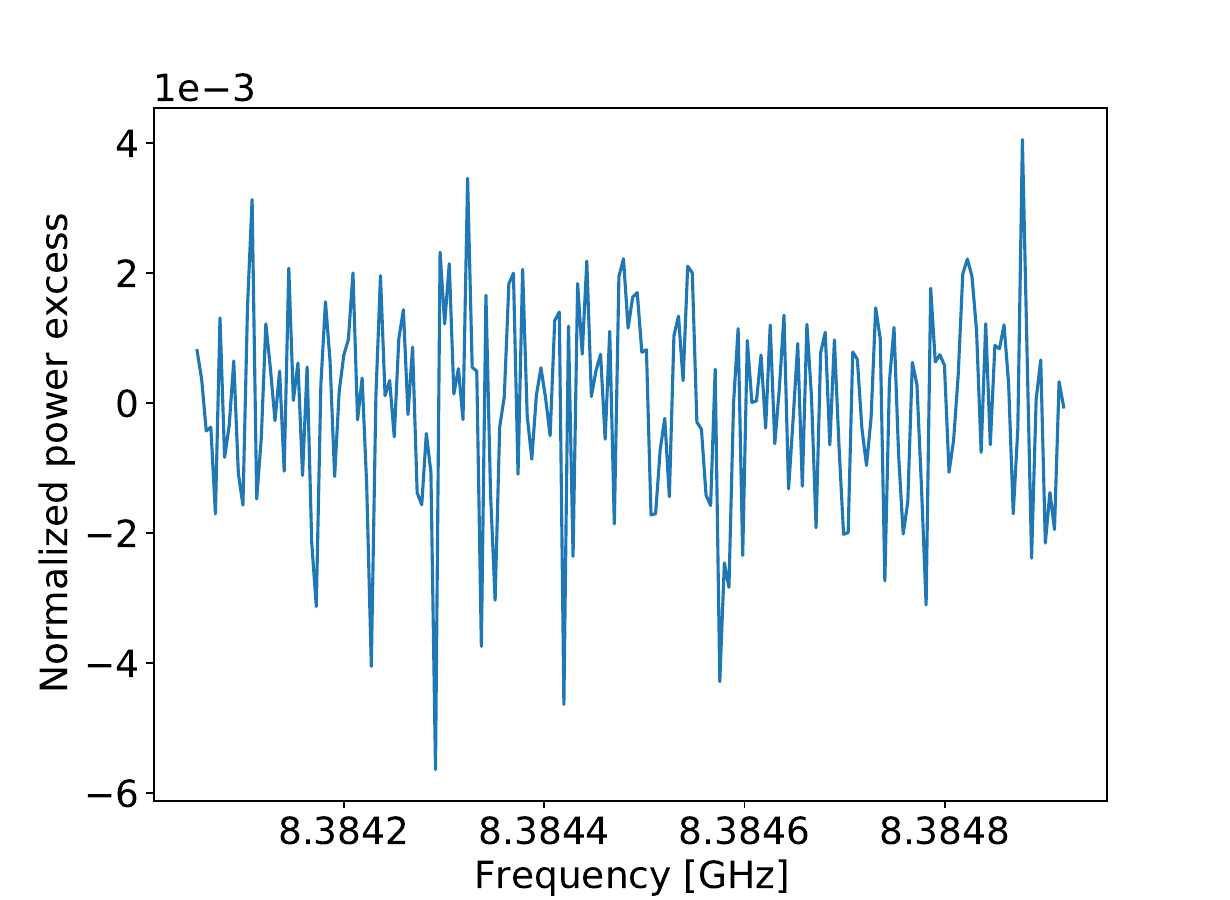}
  \caption{}
  \label{fig:Background-Removal-e}
\end{subfigure}%
\begin{subfigure}{.49\textwidth}
  \centering
  \includegraphics[width=.8\linewidth]{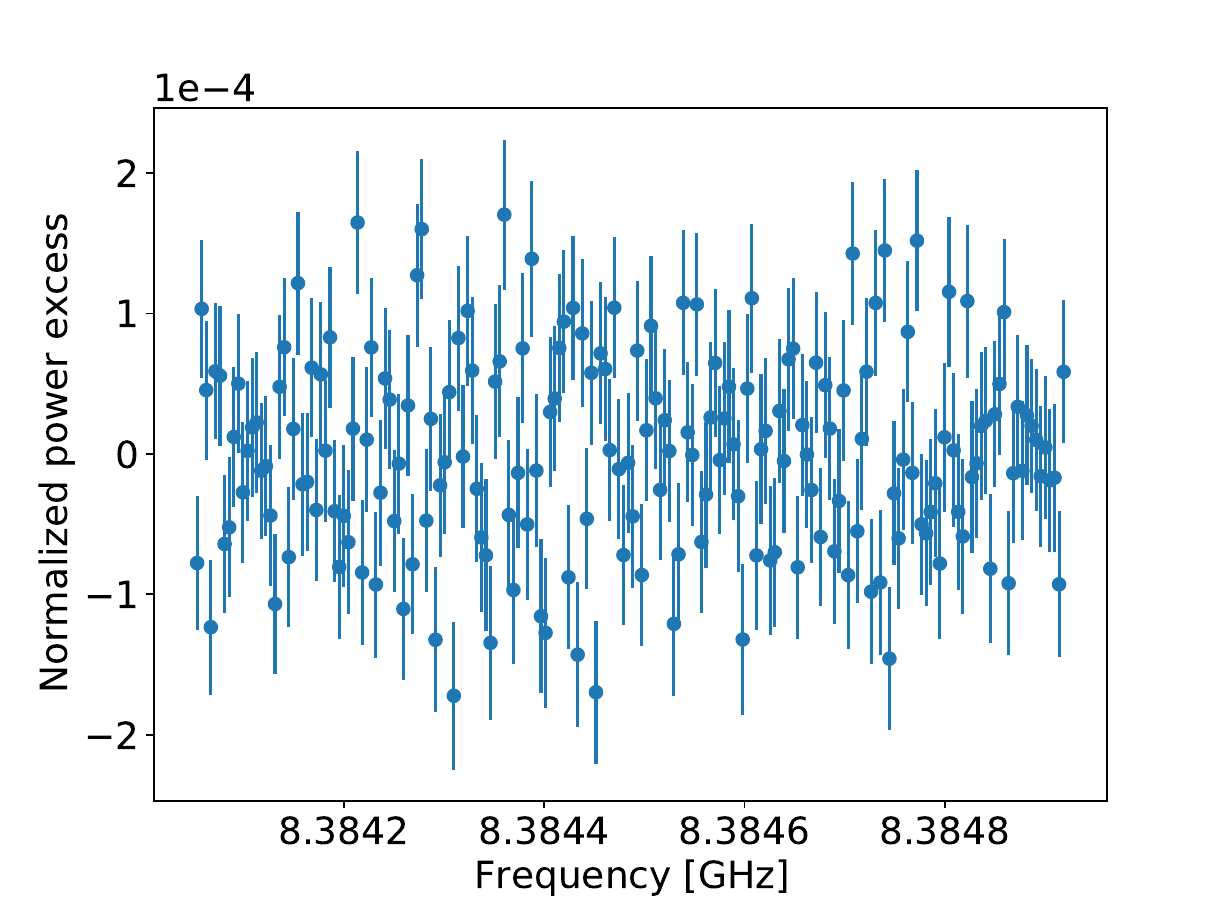}
  \caption{}
  \label{fig:Background-Removal-f}
\end{subfigure}
\caption{Removal procedure of the electronic background. (\textbf{a}): The initial raw spectra taken by the DAQ are combinations of electronic background, the cavity resonance peak, white noise, and a possible axion signal. The electronic background and cavity resonance were removed by (\textbf{b}): dividing the spectra at different LO frequencies and (\textbf{c},\textbf{d}): applying two SG filters. (\textbf{e}): The unit-less normalized spectra were combined into a grand unified spectrum. (\textbf{f}): The magnet-on and off grand unified spectra were subtracted to remove the systematic residual structure and create the final spectrum. Figures taken from \cite{Alvarez2021}.}
\label{fig:Background-Removal}
\end{figure}

These 90s spectra (see Figure \ref{fig:Background-Removal}a) are combinations of the electronic background, the cavity resonance peak structure, thermal noise, and any possible axion signal. The unwanted electronic background and the cavity resonance structure were removed following these steps:

\begin{enumerate}
    \item Dividing each 90s spectrum taken at $l_1$ by the average spectrum of $l_{2-\text{on}}$. This was a first-order correction to the electronic background introduced by the DAQ (see \mbox{Figure \ref{fig:Background-Removal}b}).
    \item Limiting the frequency region of the analysis to a range of $\sim$ 0.87 MHz around the resonance peak (see Figure \ref{fig:Background-Removal}b). This range covered more than the full width at half maximum of the Lorentzian peak.
    \item Applying a Savitzky-Golay (SG) filter \cite{savitzky64, SG-Filters-2} to the average spectrum of the 90s spectra taken at $l_1$. The filter produced a fit labeled SG-fit (see Figure \ref{fig:Background-Removal}c).
    \item Normalized spectra were created by dividing each 90s spectrum by the SG-fit (see Figure \ref{fig:Background-Removal}d).
    \item Due to drifts in the receiver chain, the normalized spectra still had unwanted structure remaining within the spectra. A second SG-fit (SG$^*$) was created for each normalized 90s spectrum. The spectra were divided by the SG$^*$-fits, and the result was subtracted by 1 to create unit-less normalized power spectra (see Figure \ref{fig:Background-Removal}e).
    \item The unit-less normalized power spectra were combined into a grand unified spectrum.
    \item The previous steps were repeated for the magnet-off data sets.
    \item The final spectrum was the difference of the magnet-on and magnet-off grand unified spectra (see Figure \ref{fig:Background-Removal}f).
\end{enumerate}

In step 1, by changing the LO frequency, the resonance peak was displaced in the IF frequencies introduced in (\ref{IF}). However, the intrinsic background structure introduced by the DAQ remained qualitatively similar. The SG-fit of step 3 was constructed to remove large frequency scale features, such as the resonance peak; and the SG$^*$-fits of step 5 removed time scale variations, such as the one produced by gain drifts.

Spectra produced after step 5 should be a combination of thermal noise and a possible axion signal. Thermal noise follows a Gaussian distribution, and the expected fluctuation of these spectra is given by:
\begin{equation}
    \label{eq:Noise-fluctuation}
    \sigma = \frac{1}{\sqrt{t\cdot \Delta \nu}},
\end{equation}
where $\Delta \nu$ is the resolution bandwidth of the DAQ (4577 Hz) and $t$ is the integrated time of the power spectra. For $t = 90$ s, the theoretical $\sigma = 0.001554$. The expected values of the pull distribution of a Gaussian distributed value are a mean value $\mu = 0$ and a width of $\sigma = 1$. The histogram of the pull distribution of the unit-less normalized power spectra was created, and a Gaussian distribution was fitted to it. The fit gave a mean value $\mu = 0.0002 \pm 0.0009$ and a width of $\sigma = 1.0048 \pm 0.0009$, which proved that at the 90 s fluctuation level, the background structure was removed. After combining the 103 hours of magnet-on together, the pull distribution of the histogram of the grand unified spectrum showed an unexpected non-Gaussian structure with $\mu = 0.0 \pm 0.2$ and $\sigma = 1.7 \pm 0.2$.

This structure was identified as a systematic residual, which also appeared in the magnet-off data. Step 8 was done to remove this structure without affecting a possible axion signal. The fitted Gaussian function to the final spectrum yielded $\mu = 0.00 \pm 0.08$ and $\sigma = 1.07 \pm 0.08$. An axion search was done to this spectrum.

To perform the search, the axion line shape was created using the velocity distribution based of the standard isothermal spherical halo model \cite{Jimenez:2002vy}. The two SG-fits were applied to the line shape to replicate the effects of the background removal, steps 3 to 5. A signal attenuation of 20 $\%$ and a distortion of the line shape were the results of applying the SG-filters. The resulting distorted line shape (which had a range of 16 frequency bins) was fitted through the 190 frequency bins of the final spectrum. The amplitude $A$ given by the fit function represented the axion power deposited at each frequency. After applying the look-elsewhere effect \cite{Lista:2017jsy}, the highest $A$ excess yielded a global significance of $\sigma = 3.05$. Thus no significant signal above statistical fluctuations was observed.

The Bayesian method \cite{Lista:2017jsy} was used to compute the upper limit (UL). The fact that a possible axion signal would appear as a positive $A$ value on the fit function was used as the prior for the posterior function. The likelihood function of the normalized power excess values of the final spectrum can be expressed with a $\chi^2$ distribution because the variable was Gaussian distributed. The UL of $A$ ($A_{UL}$) was computed integrating over the posterior function, as shown in the following equation:
\begin{equation}
    \label{eq:Upper-Limit-p}
    \frac{1}{N}  \int_0^{A_{\text{UL}}} \text{e}^{-(\chi^2/2)} dA = 1 - \alpha,
\end{equation}
where 1 $-$ $\alpha$ is the credibility level (CL) of the UL. For these results a 95 $\%$ CL was used. The unit-less $A_{UL}$ can be converted back to power units with:
\begin{equation}
    \label{eq:AL-->Power}
    P_d = A_{\text{UL}} \cdot P_N = A_{\text{UL}} k_b T_{\text{sys}} \Delta \nu,
\end{equation}
and with (\ref{Pd}) one can compute the exclusion limit to axion--photon coupling $g_{a\gamma}$ of this measurement.

To evaluate (\ref{Pd}) and (\ref{eq:AL-->Power}), the external parameters of the system were measured. For the $T_{\text{sys}}$, the Y-method \cite{Y-method} was followed: the loaded Q-value was computed using $Q_l = f/\Delta f$ where $f$ was the resonant frequency of the cavity's first mode and $\Delta f$ was the resonance width. The cavity coupling was measured using the reflection coefficient $S_{11}$\endnote{To measure the cavity coupling, an RF switch was installed after the data taking. Bypassing the LNA was thus not possible during data taking. The $Q_l$ measured after the installation of the switch was similar to the one computed during the data-taking period. From this result it can be assumed that the  coupling measured before and after the data-taking period were the same.}. The geometric factor and volume were taken from simulations. The resolution bandwidth was determined by the DAQ specifications described in Section \ref{section:Data Acquisition system}; the standard dark matter density value used by other axion haloscope experiments was used to enable comparisons with other results. The magnetic field was provided by the CAST collaboration. Finally, an attenuation factor ($\eta$) had to be included to account for the losses between the cavity and the LNA. 

\begin{figure}[H]
\includegraphics[width=0.9 \textwidth]{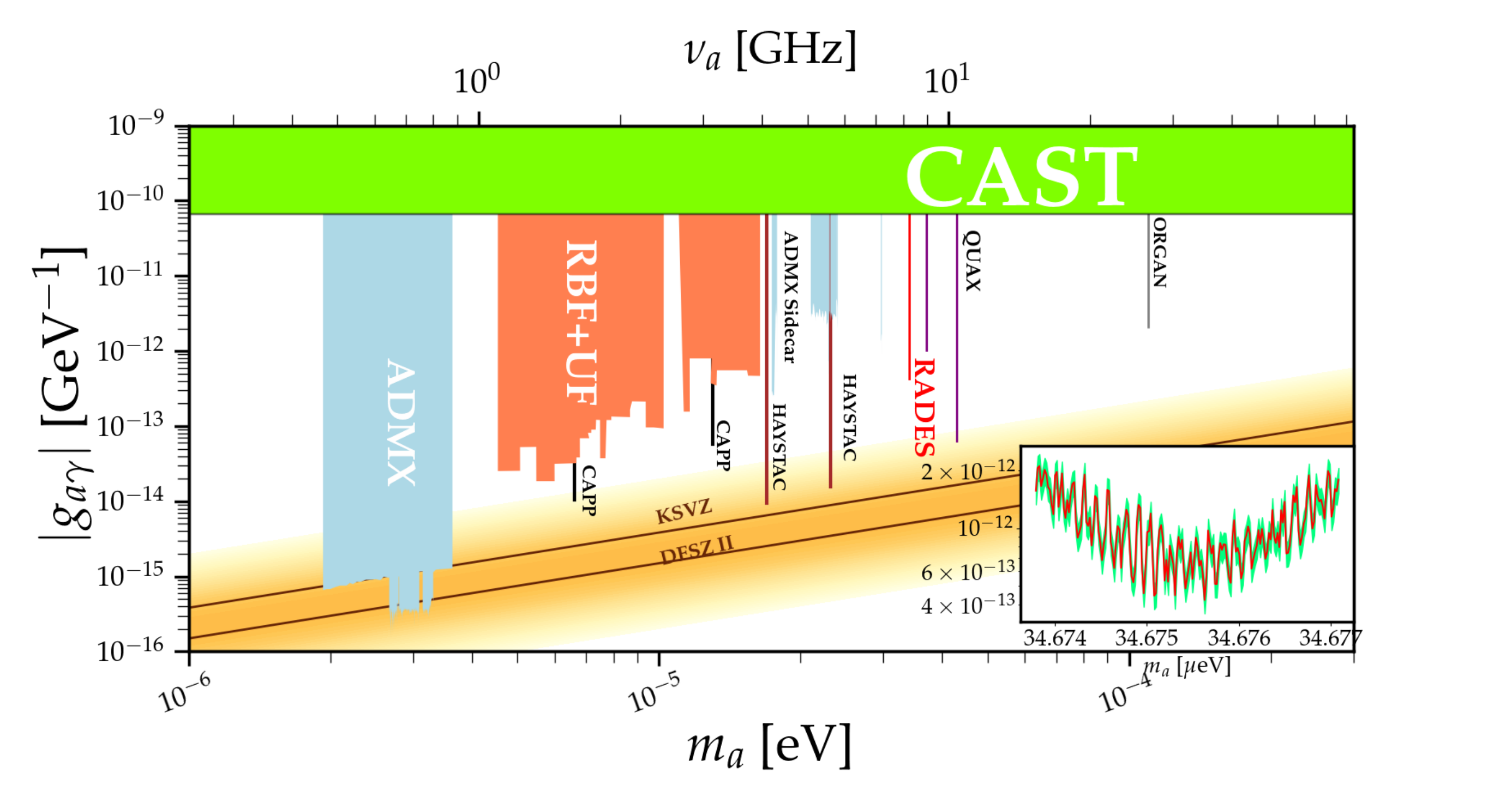}
	\caption{\label{fig:Exclusion-Limit} 
Axion--photon coupling vs. axion mass phase-space. In red is the CAST-RADES axion--photon coupling exclusion limit compared to other haloscope results and the CAST solar axion results. Inset: Zoom-in of the parameter range probed in the first CAST-RADES results (\mbox{$34.6738~\upmu$eV$ < m_a < 34.6771~\upmu$eV}), where the green region represents the uncertainty of the measurement. Figure taken from \cite{Alvarez2021}.}
\end{figure}

Table \ref{tab:Axion-paramters} lists the values for all the parameters needed to compute $g_{a\gamma}$. An exclusion limit with a 95\% credibility level on the axion--photon coupling constant of \mbox{g$_{a\gamma}\gtrsim 4\times10^{-13}$~GeV$^{-1}$} over a mass range of 34.6738 $\upmu$eV $< m_a <$ 34.6771 $\upmu$eV was achieved. This constitutes a significant improvement over the current strongest limit set by CAST at this mass. A detailed description of the first results can be found in \cite{Alvarez2021}. Figure \ref{fig:Exclusion-Limit} compares the CAST-RADES results with other haloscopes experiments and CAST's solar axion results. 

\vspace{-6pt}
\begin{table}[H]
\footnotesize
\caption{\label{tab:Axion-paramters} Parameter used to compute the exclusion limit of the axion--photon coupling.} 
\setlength{\tabcolsep}{26.2mm}
\begin{tabular}{cc}
\toprule
\textbf{Parameter} & \textbf{Value} \\ \midrule
$\Delta \nu$ & 4577 Hz \\ 
$T_{\text{sys}}$ & (7.8 $\pm$ 2.0) K \\ 
$Q_l$ & $11009 \pm 483$ \\ 
$\kappa$ & 0.33 $\pm$ 0.05 \\ 
$B_e$ & (8.8 $\pm$ 0.0088) T \\ 
$\rho_a$ & 0.45 GeVcm$^{-3}$ \\ 
$C$ & 0.65 \\ 
Volume & 0.03 l \\ 
$\eta$ & 0.83 \\ \bottomrule
\end{tabular}
\end{table}

\vspace{-15pt}

\section{Envisaged Work}

After the encouraging results of these first few years, the collaboration is currently and will be in the next years focusing on increasing the sensitivity of the experiment, on increasing the mass range by means of tuning mechanisms or materials, on developing new haloscopes for other mass regions, and on developing computational techniques which can efficiently obtain an accurate solution for the electromagnetic field induced by the axion--photon interaction. Next, these future goals are briefly described.

\subsection{Increasing the Experiment's Sensitivity}

As said before, the experiment's sensitivity, in terms of the haloscope parameters, depends on its volume, its quality factor, and its coupling to the line which extracts the signal generated by the axion--photon conversion.

The volume can be greatly increased if many cavities are used. Since our operation mode is the $TE_{101}$, the detection frequency is independent of the cavity height, depending only on its width and length. In this way, the haloscope height can be increased until it reaches the magnet bore diameter. However, mode clustering around the $TE_{101}$ mode will appear due to the new resonant $TE_{1n1}$ and $TM_{1n1}$ modes (with $n=1,2,3,..$) for very large numbers of cavities. We are now studying the proper combination of different ports of the haloscope that can alleviate this clustering and make possible this increase in volume without distorting the mode $TE_{101}$ resonance.

Moreover, the multi-cavity concept can be used in other dimension, creating 2D multi-cavities, which not only increase their volume and exploit the magnet's bore room, but also are able to introduce transmission zeros into the haloscope response for the rejection of unwanted cavity modes.

Regarding the improvement of the quality factor, a considerable increase in the experiment sensitivity is expected from the use of superconductive materials, and we recently manufactured a few cavities with the specific purpose of verifying the potential of this technology. The design of these new cavities is specifically tailored to allow soldering high-temperature superconductor (HTS) coated tapes of commercial production on the cavity surface. These tapes are optimized for magnetic flux pinning and are thus ideal for operation in a strong magnetic field. Measurement of a first prototype covered with such HTS tapes has shown a more than 50\% improvement in quality factor compared to copper, when tested at 4.2 K in a magnetic field up to 11.5 T \cite{Golm:2021ooj}. Further improvements in the soldering process have already been identified and are under testing, which should allow reaching even higher performance.

Finally, as remarked before, the sensitivity of the coaxial cavity coupling with the position, dimensions, and shape of the probe makes it advisable to develop mechanical systems for modifying this coupling in situ and get the desired critical coupling. This moving coupling system is mandatory when a frequency range is explored by tuning the haloscope properly. The collaboration is currently designing the moving coupling system with nanopositioners. For the tuning system, both mechanical and electromagnetic mechanisms, based on ferroelectric and ferromagnetic materials, are being studied.

\subsection{Haloscopes for Other Frequencies}

As described above, the first four years of the RADES collaboration have been focused on developments for the 34--35 $\upmu$eV mass range (8.2--10.6 GHz frequency range). However, the possibility of operation in future magnets has meant that part of the work of the collaboration is nowadays devoted to the design of haloscopes in two far regions \mbox{regarding this.}

First, the future BabyIAXO magnet \cite{IAXO2021,Dafni2021}, a first step of the IAXO \mbox{experiment \cite{Armengaud2019}}, with two bores of 70 cm in diameter and 10 meter in length, allows one to test new bi-dimensional multi-cavity concepts in order to increase the volume in other dimensions. Additionally, even more importantly, this magnet has enough volume to even host single cavities for searching in the 1--2 $\upmu$eV mass range (240--480 MHz, the lower limit of UHF band). These masses are just below the region explored by ADMX currently. With a 2 T magnetic field, BabyIAXO will be able to not only develop helioscope searches, but also haloscope experiments with a figure of merit ($B^2V$) of 15.5 T$^2$m$^3$, a great leap from 0.014 T$^2$m$^3$ in CAST. Currently, RADES is designing these new cavities that profit from the large volume in BabyIAXO.

Second, the Canfranc Underground Laboratory (LSC) will have at the end of 2022 a dilution cryostat working at 10 mK with a 10 T solenoid magnet, and this opens up possibilities of continuing with the development of multicavities in X band, adapted to a solenoid magnetic field direction, with an improved noise temperature, and additionally, working on haloscopes in a theoretically well motivated region for searching the axion, the 310--455 $\upmu$eV (75--110 GHz, W band). Both experiments will need detection devices that exploit the ultra-low temperatures and go beyond the standard quantum limit, such as Josephson parametric amplifiers (JPA), superconducting qubit-based single photon counters, or for higher frequencies, kinetic inductor devices (KID).

\subsection{Developments in Numerical Simulation: Application of the BI-RME 3D Method to Axion-Photon Coupling in Resonant Cavities}

The Boundary Integral Resonant Mode Expansion 3D (BI-RME 3D) \cite{birme3Dcavities,conciauro} is a full-wave modal method based on classical numerical techniques used for electromagnetic analysis. The BI-RME 3D method, developed during the eighties and nineties, provides an exact formulation which allows one to calculate the electromagnetic field existing in a lossy microwave resonator in terms of the electric and magnetic charge and current densities. 

By applying the BI-RME 3D method to the axion field, it is possible to obtain the complete information of the extracted signal from the cavity while taking into account not only the cavity shape, but also metallic and dielectric materials. This knowledge could be useful for further experiments or analysis in which the axion phase plays an important role, in particular when several waveguide ports are included in the analysis, avoiding the typical Cauchy--Lorentz approximation \cite{Younggeun2019}.

In summary, we envisage a new approach to calculate the extracted and dissipated power using a semi-analytical solution through the BI-RME 3D method, obtaining the frequency spectrum of the electromagnetic axion field generated inside the cavity. This rigorous broadband result could be of special interest in the cases where neighboring modes are quite close to the axion resonant mode and might interfere.

\section{Conclusions}

The RADES collaboration joined in 2016 the international quest for dark matter axion detection, and during the last five years has developed a new kind of resonant haloscope based on the {multi-cavity concept. The multi-cavity technology increases the volume without introducing clustering of higher resonant modes near the signal mode, and thus increases the detection power and sensitivity to the axion--photon coupling constant. The multi-cavity concept was manufactured and tested for both} inductive and alternated internal coupling, showing an improvement in the isolation of the axion mode in the second case. {Based on data collected by RADES between the 30th of October and the 20th of December of 2018, the hypothesis of the axion--photon coupling constant of \mbox{$g_{a\gamma}\gtrsim4\times10^{-13}$ GeV$^{-1}$} is rejected at a significance of 95$\%$ over a mass range of 34.6738 $\upmu$eV $< m_a <$ 34.6771 $\upmu$eV.} The future is full of new research possibilities for increasing the sensitivity of the experiment and for designing new haloscopes in lower (1--2 $\upmu$eV) and higher (310--455 $\upmu$eV) mass ranges, or for developing new computational techniques for obtaining rigorous electromagnetic solutions for the axion--photon coupling in resonant cavities.

\vspace{6pt}
\authorcontributions{This review is a brief description of the work developed by the RADES researchers in the last five years. Alejandro D\'iaz-Morcillo, Antonio J. Lozano-Guerrero, Benito Gimeno, Alejandro \'Alvarez Melc\'on, José Mar\'ia Garc\'ia Barcel\'o and Pablo Navarro have contributed to the haloscopes’ design, manufacturing and characterization, and the development of a computational technique, based on BIRME-3D, to model the axion-photon interaction in a resonant cavity; Sergio Arguedas Cuendis, Juan Daniel Gallego and Jordi Miralda-Escud\'e to the data acquisition system and data analysis; Igor Garc\'ia Irastorza, Babette D\"obrich, Carlos Pe\~na Garay, Cristian Cogollos and Chloe Malbrunot to the theoretical principles of the axion – photon interaction, the experiment definition and management at CERN; Javier Redondo to the mathematical formalism for the multi-cavity concept; Jessica Golm and Sergio Calatroni to the characterization of high thermal superconducting cavities. and Walter Wuensch to haloscopes’ design, data analysis and experiment management.} 

\funding{This work has been funded by the Spanish Agencia Estatal de Investigación (AEI) and Fondo Europeo de Desarrollo Regional (FEDER) under projects FPA-2016-76978-C3-2-P (supported by the grant FPI BES-2017-079787) and PID2019-108122GB-C33, and was supported by the CERN Doctoral Studentship programme. The research leading to these results has received funding from the European Research Council  under grant ERC-2018-StG-802836 (AxScale project). IGI acknowledges support from the European Research Council (ERC) under grant ERC-2017-AdG-788781 \mbox{(IAXO+ project).}}
\conflictsofinterest{The authors declare no conflict of interest.} 

\begin{adjustwidth}{-\extralength}{0cm}
\printendnotes[custom]
\reftitle{References} 




\end{adjustwidth}
\end{document}